\documentclass[12pt, draftcls, one column, journal]{IEEEtran}
\usepackage{amsmath,amssymb,epsfig,bbm,amsbsy,cite}
\usepackage{color}
\usepackage{amsmath}
\usepackage{amssymb}

\begin{document}
\title{Robust Adaptive Beamforming Based on Steering Vector Estimation
via Semidefinite Programming Relaxation}

\author{Arash Khabbazibasmenj,~\IEEEmembership{Student Member,~IEEE,}
        Sergiy A. Vorobyov,~\IEEEmembership{Senior Member,~IEEE,}
        Aboulnasr Hassanien,~\IEEEmembership{Member,~IEEE}

\thanks{This work is supported in parts by the Natural Science and
Engineering Research Council (NSERC) of Canada and the Alberta
Ingenuity Foundation, Alberta, Canada.

The authors are with the Department of Electrical and Computer
Engineering, University of Alberta, 9107-116 St., Edmonton,
Alberta, T6G~2V4 Canada. Emails: {\tt \{khabazi, vorobyov,
hassanie \}@ece.ualberta.ca}

{\bf Corresponding author:} Sergiy A.~Vorobyov, Dept. Elect. and
Comp. Eng., University of Alberta, 9107-116 St., Edmonton,
Alberta, T6G 2V4, Canada; Phone: +1 780 492 9702, Fax: +1 780 492
1811. Email: {\tt vorobyov@ece.ualberta.ca}. } }
\vspace{-1.5cm}
\maketitle
\vspace{-1.5cm}

\begin{abstract}
We develop a new approach to robust adaptive beamforming in the
presence of signal steering vector errors. Since the signal
steering vector is known imprecisely, its presumed (prior) value
is used to find a more accurate estimate of the actual steering
vector, which then is used for obtaining the optimal beamforming
weight vector. The objective for finding such an estimate of the
actual signal steering vector is the maximization of the
beamformer output power, while the constraints are the
normalization condition and the requirement that the estimate of
the steering vector does not converge to an interference steering
vector. Our objective and constraints are free of any design
parameters of non-unique choice. The resulting optimization
problem is a non-convex quadratically constrained quadratic
program, which is NP hard in general. However, for our problem we
show that an efficient solution can be found using the
semi-definite relaxation technique. Moreover, the strong duality
holds for the proposed problem and can also be used for finding
the optimal solution efficiently and at low complexity. In some
special cases, the solution can be even found in closed-form. Our
simulation results demonstrate the superiority of the proposed
method over other previously developed robust adaptive beamforming
methods for several frequently encountered types of signal
steering vector errors.
\end{abstract}

\vspace{-0.4cm}

\maketitle

\IEEEdisplaynotcompsoctitleabstractindextext
\IEEEpeerreviewmaketitle

\section{Introduction}
Robust adaptive beamforming design has been an intensive research
topic over several decades due to, on one hand, its ubiquitous
applicability in wireless communications, radar, sonar, microphone
array speech processing, radio astronomy, medical imaging, and so
on; and on the other hand, because of the challenges related to
the practical applications manifesting themselves in the
robustness requirements for adaptive beamformers. The main causes
of performance degradation in adaptive beamforming are small
sample size and imprecise knowledge of the desired signal steering
vector in the situation when the desired signal components are
present in the training data. The traditional design approaches to
adaptive beamforming \cite{Trees}-\cite{Hung} do not provide
sufficient robustness and are not applicable in such situations.
Thus, various robust adaptive beamforming techniques gained a
significant popularity due to their practical importance
\cite{Gershman2}. The most popular conventional robust adaptive
beamforming approaches are the diagonal loading technique
\cite{Cox}, \cite{Abramovich} and the eigenspace-based beamforming
technique \cite{Feldman}, \cite{Chang}. However, it is not clear
in the former approach how to obtain the optimal value of the
diagonal loading factor, whereas the eigenspace-based beamforming
is known to suffer from the so-called subspace swap phenomenon at
low signal-to-noise ratios (SNRs) \cite{Scharf} and requires exact
knowledge of the signal-plus-interference subspace dimension.

In the last decade, several approaches to robust adaptive
beamforming based on a rigorous modeling of the steering vector
mismatches have been developed. For the case when the mismatch of
the signal steering vector is modeled as deterministic unknown
norm bounded vector, the so-called worst-case-based adaptive
beamforming design has been proposed in \cite{Vorobyov1},
\cite{Vorobyov2}. The relationship of the worst-case-based design
to the diagonal loading principle with adaptive diagonal loading
factor has been explored in \cite{Vorobyov1}, \cite{LiStoica} as
well as the generalization to the ellipsoidal steering vector
uncertainty set has been developed in\cite{Lorenz}. The
uncertainty of second order statistics of the desired signal have
been also considered in \cite{Shahbaz}. If the signal steering
vector mismatch is modeled as random unknown with known Gaussian
or unknown distribution, the corresponding optimization problem
belongs to the class of stochastic programming problems, and the
corresponding probabilistically-constrained robust adaptive
beamformer has been developed in \cite{Vorobyov7},
\cite{Vorobyov4}. The relationship between the worst-case-based
and the probabilistically-constrained robust adaptive beamformers
has also been shown in \cite{Vorobyov4}, \cite{Vorobyov3}.
However, some design parameters  such as the norm of the signal
steering vector mismatch in the worst-case-based design or the
acceptable beamforming outage probability in the
probabilistically-constrained design are assumed to be known in
the aforementioned techniques.

Another approach to robust adaptive beamforming which is based on
estimating actual signal steering vector based on the knowledge of
presumed steering vector has been proposed in \cite{Nasr},
\cite{Vorobyov5}. The main idea of this approach is to estimate
the signal steering vector so that the maximum of the beamformer
output power is achieved, while the convergence of the steering
vector estimate to an interference steering vector is avoided. The
latter convergence can be avoided by imposing different
constraints. The projection constraint to the space to which the
desired signal belongs is used in \cite{Vorobyov5} and the
solution based on sequential quadratic programming (SQP) is
developed. The advantage of the method of \cite{Vorobyov5} is that
no design parameters of non-unique choice are required. However,
the disadvantage is that the complexity of SQP is rather high,
which makes the method less attractive for practical use. It is
interesting that the robust adaptive beamforming formulation based
on desired signal steering vector estimation has been also
considered in \cite{LiStoicaWang}, but the typical for the
worst-case-based methods \cite{Vorobyov1}-\cite{Lorenz} norm bound
constraint on the steering vector mismatch has been used there in
order to guarantee that the estimate of the desired signal
steering vector does not converge to an interference steering
vector. As a result, the method of \cite{LiStoicaWang} is one of
the various implementations of the adaptive diagonal loading based
techniques, and it is not free of design parameters.

In this paper, we develop a new robust adaptive beamforming method
which is free of any design parameters of non-unique choice. This
method is based on the signal steering vector estimation via
beamformer output power maximization under the constraint on the
norm of the steering vector estimate and the requirement that the
estimate of the steering vector does not converge to an
interference steering vector. To satisfy the latter requirement,
we develop a new constraint which is a convex quadratic
constraint. Then the corresponding optimization problem is a
non-convex (due to the steering vector normalization condition)
homogeneous quadratically constrained quadratic programming (QCQP)
problem. In general, QCQP problems may not have a strong duality
property, which leads to the situation when the solution of the
corresponding semidefinite programming (SDP) relaxation-based
problem is of rank higher than one and randomization procedures
have to be used to find an approximate rank-one solution
\cite{Luo2010}-\cite{Zhang}. The probability that the so-obtained
rank-one solution coincides with the exact solution is less than
one \cite{Tom}. However, in the case of our signal steering vector
estimation problem, we show that a rank-one solution can be found
efficiently using the SDP relaxation technique. Moreover, the
strong duality holds for the proposed problem, which means that a
rank-one solution can also be found based on the solution of the
convex dual problem. Some special cases and interesting
relationships are also considered. Our simulation results
demonstrate the superiority of the proposed method over other
previously developed robust adaptive beamforming techniques.

This paper is organized as follows. Data model, beamforming
formulation, and necessary background are given in Section~II. In
Section~III, we formulate the problem of interest. A complete
analysis of the problem and its rank-one solution is given in
Section~IV. Section~V overviews some special cases and draws
interesting relations to the existing methods. Simulation results
comparing the performance of the proposed method to the existing
methods are shown in Section~VI. Finally, Section~VII presents our
conclusions.

\section{System Model and Background}
Consider a linear antenna array with $M$ omni-directional antenna
elements. The narrowband signal received by the antenna array at
the time instant $k$ can be written as
\begin{equation} \label{signal_array}
\mathbf x (k)= \mathbf s (k) + \mathbf i (k) + \mathbf n (k)
\end{equation}
where $\mathbf s (k)$, $\mathbf i (k)$, and $\mathbf n (k)$ denote
the $M \times 1$ vectors of the desired signal, interference, and
noise, respectively. The desired signal, interference, and noise
components of the received signal \eqref{signal_array} are assumed
to be statistically independent to each other. The desired signal
can be written as $\mathbf s (k) = s(k) \mathbf a $, where $s (k)$
is the desired signal waveform and $\mathbf a $ is the steering
vector associated with the desired signal.

The beamformer output at the time instant $k$ can be written as
\begin{equation} \label{output_signal}
y (k) = \mathbf w ^H \mathbf x (k)
\end{equation}
where $\mathbf w$ is the $M \times 1$ complex weight (beamforming)
vector of the antenna array and $( \cdot )^H$ stands for the
Hermitian transpose.

Assuming that the steering vector $\mathbf a $ is known precisely,
the optimal weight vector can be obtained by maximizing the
beamformer output signal-to-noise-plus-interference ratio (SINR)
\cite{Trees}
\begin{equation} \label{SINR}
{SINR}= \frac{\sigma_{\rm s}^2 | \mathbf w^H \mathbf a
|^2}{\mathbf w^H \mathbf R_{i+n} \mathbf w}
\end{equation}
where $\sigma_{\rm s}^2$ is the desired signal power, $\mathbf
R_{i+n} \triangleq E \{ (\mathbf i (k) + \mathbf n (k))(\mathbf i
(k) + \mathbf n (k))^H \}$ is the $M \times M$
interference-plus-noise covariance matrix, and $E \{ \cdot \}$
stands for the statistical expectation. Since $\mathbf R_{i+n}$ is
unknown in practice, it is substituted in \eqref{SINR} by the data
sample covariance matrix
\begin{equation} \label{datasample}
\hat{\mathbf R} = \frac{1}{K} \sum_{i=1}^K \mathbf x (i) \mathbf
x^H (i)
\end{equation}
where $K$ is the number of training data samples which also
include the desired signal component. The sample version of the
problem of maximizing \eqref{SINR} is known as the minimum
variance (MV) sample matrix inversion (SMI) beamforming and it is
based on its conversion to the mathematically equivalent problem
of minimizing the denominator of \eqref{SINR} under fixed
numerator, that is,
\begin{equation} \label{SMI}
\min_{\mathbf w} \mathbf w^H \hat{\mathbf R} \mathbf w \quad {\rm
subject \ to} \quad \mathbf w^H \mathbf a  = 1.
\end{equation}
The problem \eqref{SMI} is convex and its solution can be easily
found as ${\mathbf w}_{\rm MV-SMI} = \alpha \hat{\mathbf R}^{-1}
\mathbf a $ where $\alpha = 1 / \mathbf a ^H \hat{\mathbf R}^{-1}
\mathbf a $ \cite{Trees}.

In practice, the steering vector $\mathbf a $ is not known
precisely and only its inaccurate estimate $\mathbf p$, called
hereafter as presumed steering vector, is available. Several
rigorous approaches, which address the problem of imprecise
knowledge of the desired signal steering vector have been
developed in the last decade. First of them assumes that the
actual steering vector $\mathbf a $ can be explicitly modeled as a
sum of the presumed steering vector and a deterministic mismatch
vector $\boldsymbol \delta$, i.e., $\mathbf a  = \mathbf p +
\boldsymbol \delta$ \cite{Vorobyov1}, \cite{Vorobyov2}. Here
$\boldsymbol \delta$ is unknown but it is known that $\|
\boldsymbol \delta \| \le \varepsilon$ for some bound value
$\varepsilon$, where $\| \cdot \|$ is the Euclidian norm of a
vector. This approach has been also generalized for ellipsoidal
uncertainty case in addition to the aforementioned spherical
uncertainty \cite{Lorenz}. Assuming spherical uncertainty set for
$\boldsymbol \delta$, i.e., $\cal{A} (\varepsilon ) \triangleq \{
\mathbf a  = \mathbf p + {\boldsymbol \delta} \, | \, \|
\boldsymbol \delta \| \le \varepsilon \}$, the worst-case-based
robust adaptive beamforming aims at solving the following
optimization problem
\begin{equation} \label{WCB}
\min_{\mathbf w} \mathbf w^H \hat{\mathbf R} \mathbf w \quad {\rm
subject \ to} \quad \max_{\mathbf a  \in \cal{A} (\varepsilon) } |
\mathbf w^H \mathbf a  | \ge 1 .
\end{equation}
In turn, the problem \eqref{WCB} is equivalent to the following
convex optimization problem \cite{Vorobyov1}
\begin{equation} \label{WCBsol}
\min_{\mathbf w} \mathbf w^H \hat{\mathbf R} \mathbf w \quad {\rm
subject \ to} \quad \varepsilon \| \mathbf w \| \le \mathbf w^H
\mathbf p - 1
\end{equation}
which can be efficiently solved using second-order cone
programming (SOCP) \cite{Vorobyov1}, \cite{Sedumi} or numerical
Lagrange multiplier techniques \cite{Lorenz}, \cite{Shahbaz2}.

Another approach to robust adaptive beamforming is based on the
assumption that the vector $\boldsymbol \delta$ is random. Then
the problem \eqref{WCB} changes to
\begin{equation} \label{ProbProblem}
\min_{\mathbf w} \mathbf w^H \hat{\mathbf R} \mathbf w \quad {\rm
subject \ to} \quad {\rm Pr} \{ | \mathbf w^H \mathbf a  | \ge 1
\} \ge p_0
\end{equation}
where ${\rm Pr} \{\cdot \}$ denotes probability and $p_0$ is
preselected probability value. In the case of Gaussian distributed
$\boldsymbol \delta$ and the case when the distribution of
$\boldsymbol \delta$ is unknown and assumed to be the worst
possible, it has been shown that the problem \eqref{ProbProblem}
can be approximated by the following problem \cite{Vorobyov4},
\cite{Vorobyov3},
\begin{equation} \label{ProbProblemsol}
\min_{\mathbf w} \mathbf w^H \hat{\mathbf R} \mathbf w \quad {\rm
subject \ to} \quad \tilde \varepsilon \| \mathbf Q_\delta^{1/2}
\mathbf w \| \le \mathbf w^H \mathbf p - 1
\end{equation}
where $\mathbf Q_\delta$ is the covariance matrix of random
mismatch vector $\boldsymbol \delta$, and $\tilde \varepsilon =
\sqrt{-\ln (1 - p_0)}$ if $\boldsymbol \delta$ is Gaussian
distributed or $\tilde \varepsilon = 1 / \sqrt{1 - p_0}$ if the
distribution of $\boldsymbol \delta$ is unknown. Both problems
\eqref{WCBsol} and \eqref{ProbProblemsol} have similar SOCP
structure and can be solved efficiently.

The third approach to robust adaptive beamforming aims at
estimating the steering vector $\mathbf a $ based on the prior
given by the presumed steering vector $\mathbf p$ \cite{Nasr},
\cite{Vorobyov5}. The estimate of the steering vector $\mathbf a $
is found so that the beamformer output power is maximized while
the convergence of the estimate of $\mathbf a $ to any
interference steering vector is prohibited. Indeed, the solution
of \eqref{SMI} can be written as a function of unknown
$\boldsymbol \delta$, that is, $\mathbf w ({\boldsymbol \delta} )
= \alpha \hat{\mathbf R}^{-1} ( \mathbf p + \boldsymbol \delta )$.
Using the latter expression, the beamformer output power can be
also written as a function of $\boldsymbol \delta$ as
\begin{equation} \label{outPower}
P ( {\boldsymbol \delta} ) = \frac{1}{( \mathbf p + \boldsymbol
\delta )^H \hat{\mathbf R}^{-1} ( \mathbf p + \boldsymbol \delta
)}.
\end{equation}
Thus, such estimate of $\boldsymbol \delta$ or, equivalently, such
estimate of $\mathbf a$ that maximizes \eqref{outPower} will be
the best estimate of the actual steering vector $\mathbf a $ under
the constraints that the norm of $\hat{\mathbf a} $ equals
$\sqrt{M}$ and $\hat{\mathbf a} $ does not converge to any of the
interference steering vectors. The latter can be guaranteed by
requiring that
\begin{equation} \label{NasrConst}
\mathbf P^\bot ( \mathbf p + \hat{\boldsymbol \delta} ) = \mathbf
P^\bot \hat{\mathbf a}  = 0
\end{equation}
where $\mathbf P^\bot = \mathbf I - \mathbf U \mathbf U^H$,
$\mathbf U = [\mathbf u_1, \mathbf u_2, \hdots, \mathbf u_L]$, $
\mathbf u_l$, ${l=1}, \hdots, L$ are the $L$ dominant eigenvectors
of the matrix $\mathbf C = \int_\Theta \mathbf d (\theta ) \mathbf
d^H (\theta ) \, d \theta$, $\mathbf d(\theta)$ is the steering
vector associated with direction $\mathbf \theta$ and having the
structure defined by the antenna geometry, $\Theta$ is the angular
sector in which the desired signal is located, $\hat{\boldsymbol
\delta}$ and $\hat{\boldsymbol a}$ stand for the estimate of the
steering vector mismatch and for the estimate of the actual
steering vector, respectively and  $\mathbf I$ is the identity
matrix. The resulting optimization problem is non-convex, but has
been solved in \cite{Vorobyov5} using SQP technique. A similar
approach based on steering vector estimation has been also
recently taken in \cite{Gershman} for the case when $\mathbf a $
is partially known, for example, when array is partially
calibrated, that significantly simplifies the problem. The
following interesting relationship is also worth mentioning. If
the constraint \eqref{NasrConst} is replaced by the constraint $\|
\boldsymbol \delta \| \leq \varepsilon$ used in the
worst-case-based beamformers, the convergence to an interference
steering vector will also be avoided, but the problem becomes
equivalent to the worst-case-based robust adaptive beamforming
(see \cite{LiStoicaWang}).

Finally, it is worth mentioning that the eigenspace-based
beamformer \cite{Feldman}, \cite{Chang} is also based on
correcting/estimating the desired signal steering vector. Taking
the presumed steering vector $\mathbf p$ as a prior, the
eigenspace-based beamformer finds and uses the projection of
$\mathbf p$ onto the sample signal-plus-interference subspace as a
corrected estimate of the steering vector. The eigendecomposition
of \eqref{datasample} yields
\begin{equation}
\hat{\mathbf R} = \mathbf E \boldsymbol \Lambda \mathbf E^H +
\mathbf G \boldsymbol \Gamma \mathbf G^H
\end{equation}
where the $M \times J+1$ matrix $\mathbf E$ and $M \times M-J-1$
matrix $\mathbf G$ contain the signal-plus-interference subspace
eigenvectors of $\hat{\mathbf R}$ and the noise subspace
eigenvectors, respectively, while the $J+1 \times J+1$ matrix
$\boldsymbol \Lambda$ and $M-J-1 \times M-J-1$ matrix $\boldsymbol
\Gamma$ contain the eigenvalues corresponding to $\mathbf E$ and
$\mathbf G$, respectively. Here, $J$ is the number of interfering
signals. Then the eigenspace-based beamformer is given by

\begin{equation} \label{eigenbeam}
\mathbf w_{\rm eig} = \hat{\mathbf R}^{-1} \hat{\mathbf a} =
\hat{\mathbf R}^{-1} \mathbf E \mathbf E^H \mathbf p = \mathbf E
\boldsymbol \Lambda^{-1} \mathbf E^H \mathbf p
\end{equation}
where $\hat{\mathbf a}  = \mathbf E \mathbf E^H \mathbf p$ is the
projection of the presumed steering vector $\mathbf p$ onto the
sample signal-plus-interference subspace and $\mathbf E \mathbf
E^H$ is the corresponding projection matrix. As compared to the
beamformer of \cite{Vorobyov5} based on the estimation of steering
vector, the eigenspace-based beamformer may suffer from a high
probability of subspace swaps as well as incorrect estimation of
the signal-plus-interference subspace dimension.

\section{New beamforming Problem Formulation}
The problem of maximizing the output power \eqref{outPower} is
equivalent to the problem of minimizing the denominator of
\eqref{outPower}. The obvious constraint that must be imposed on
the estimate $\hat{\mathbf a} $, is that the norm of $\hat{\mathbf
a}$ must be equal to $\sqrt{M}$, i.e., $\| \hat{\mathbf a}  \| =
\sqrt{M}$. This normalization condition, however, does not protect
the estimate $\hat{\mathbf a} $ from possible convergence to an
interference steering vector. In order to avoid such convergence,
we assume that the desired source is located in the angular sector
$\Theta = [ \theta_{\min}, \, \theta_{\max} ]$ which can be
obtained, for example, using low resolution direction finding
methods. The angular sector $\Theta$ is assumed to be
distinguishable from general locations of all interfering signals.
In turns, the sector $\tilde{\Theta}$ denotes the complement of
the sector $\Theta$, i.e., combines all the directions which lie
outside of $\Theta$. Let us define the $M \times M $ matrix
$\tilde{\mathbf C}$ as $\tilde{\mathbf C} = \int_{\tilde{\Theta}}
{\mathbf d (\theta ) \mathbf d^H (\theta ) \, d \theta}$. Then
constraint
\begin{equation} \label{ConstMain}
\hat{\mathbf a} ^H \tilde{\mathbf C} \hat{\mathbf a}  \leq
\Delta_0
\end{equation}
for a uniquely selected value  $\Delta_0$ (see Example~1 below),
will force the estimate $\hat{\mathbf a} $ not to converge to any
interference steering vector with the directions within the
angular sector $\tilde{\Theta}$. To illustrate how the constraint
\eqref{ConstMain} works, let us consider the following example.

\begin{figure}[p]
\begin{center}
\includegraphics[scale=.3]{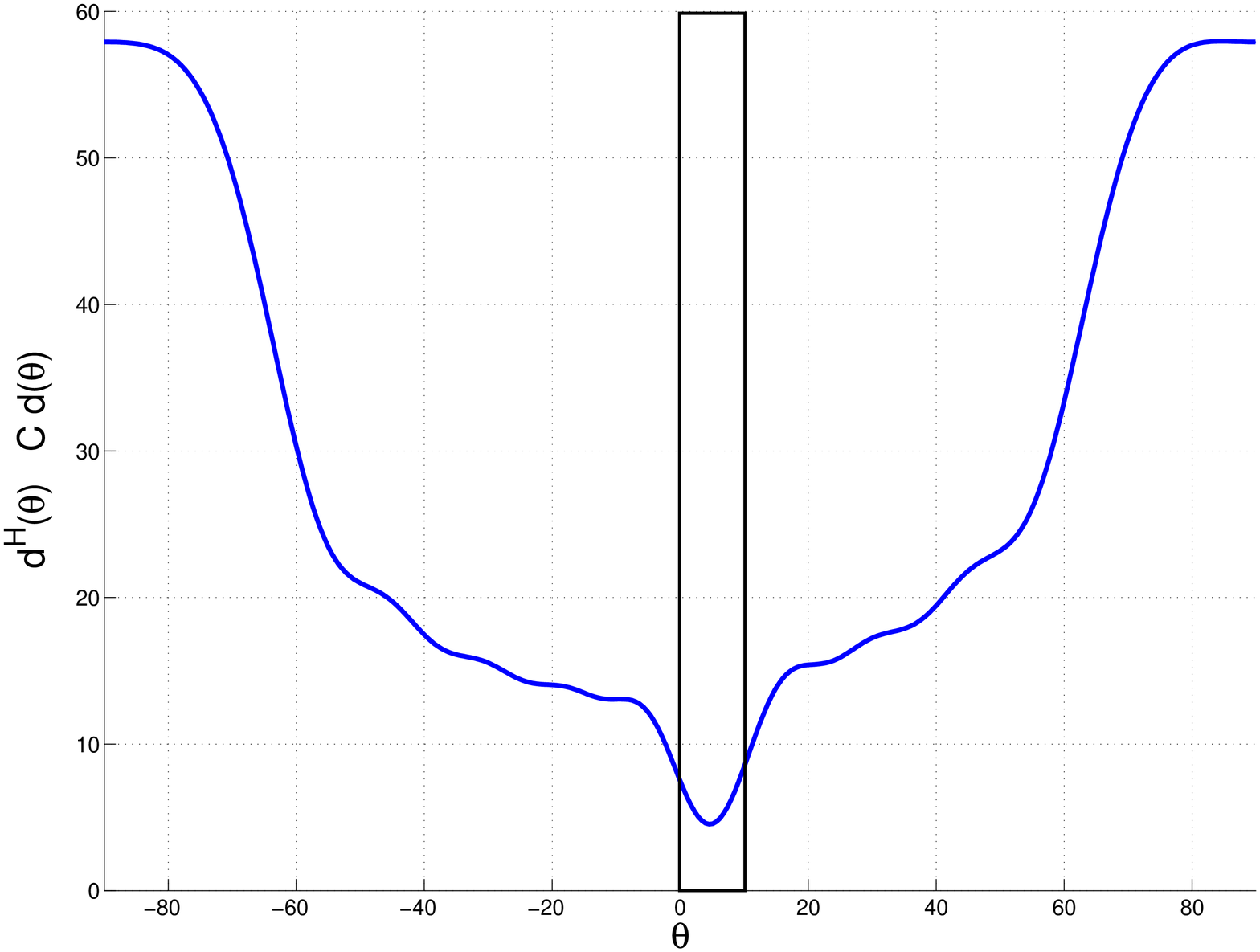}
\caption{The term $\mathbf d^H (\theta)\tilde{\mathbf C} \mathbf
d(\theta)$ in the constraint \eqref{ConstMain} versus different
angles.} \label{column_space_interpretation}
\end{center}
\end{figure}

{\bf Example~1:} Consider uniform linear array (ULA) of $10$
omni-directional antenna elements spaced half wavelength apart
from each other. Let the range of the desired signal angular
locations be $\Theta = [0^ \circ, \, 10^\circ]$.
Fig.~\ref{column_space_interpretation} depicts the values of the
quadratic term $\mathbf d^H (\theta) \tilde{\mathbf C} \mathbf
d(\theta)$ for different angles. The rectangular bar in the figure
marks the directions within the angular sector $\Theta$. It can be
observed from this figure that the term $\mathbf d^H (\theta)
\tilde{\mathbf C} \mathbf d (\theta)$ takes the smallest values
within the angular sector $\Theta$, where the desired signal is
located, and increases outside of this sector. Therefore, if
$\Delta_0$ is selected to be equal to the maximum value of the
term $\mathbf d^H (\theta ) \tilde{\mathbf C} \mathbf d (\theta )$
within the angular sector of the desired signal $\Theta$, the
constraint \eqref{ConstMain} will guarantee that the estimate of
the desired signal steering vector will not converge to any
interference steering vectors. Note also that the constraint
\eqref{ConstMain} is an alternative to the constraint
\eqref{NasrConst} used in \cite{Vorobyov5}. However, the
constraint \eqref{NasrConst} may result in the noise power
magnification at low SNRs (see \cite{Vorobyov5}), while the
constraint \eqref{ConstMain} helps to alleviate the effect of the
noise power magnifying at low SNRs by not collecting the noise
power from the continuum of the out-of-sector directions $\tilde
\Theta$.


Taking into account the normalization constraint and the
constraint \eqref{ConstMain}, the problem of estimating the
desired signal steering vector based on the knowledge of the prior
$\mathbf p$ can be formulated as the following optimization
problem
\begin{eqnarray}
\min_{ \hat{\mathbf a}  } && \hat{\mathbf a} ^H \hat{\mathbf
R}^{-1} \hat{\mathbf a}
\label{Converted_to_real_1_1} \\
{\rm subject \ to} && \left\| \hat{\mathbf a}  \right\| = \sqrt{M}
\label{Converted_to_real_1_2} \\
&& \hat{\mathbf a} ^H  \tilde{\mathbf C} \hat{\mathbf a} \leq
\Delta_0 . \label{Converted_to_real_1_3}
\end{eqnarray}
where the prior $\mathbf p$ is used only for selecting the sector $\Theta$.
Due to the equality constraint (\ref{Converted_to_real_1_2}),
which is a non-convex one, the QCQP problem of type
(\ref{Converted_to_real_1_1})--(\ref{Converted_to_real_1_3}) is
non-convex and an NP-hard in general. However, as we show
in the following section, an exact and simple solution
specifically for the problem
(\ref{Converted_to_real_1_1})--(\ref{Converted_to_real_1_3}) can
be found using the SDP relaxation technique and the strong duality theory.

\section{Steering Vector Estimation via Semi-Definite Programming
Relaxation} QCQP problems of type
(\ref{Converted_to_real_1_1})--(\ref{Converted_to_real_1_3}) can
be solved using SDP relaxation technique. The first step is to
make sure that the problem
(\ref{Converted_to_real_1_1})--(\ref{Converted_to_real_1_3}) is
feasible. Fortunately, it can be easily verified that
(\ref{Converted_to_real_1_1})--(\ref{Converted_to_real_1_3}) is
feasible if and only if $\Delta_0 / M$ is greater than or equal to
the smallest eigenvalue of the matrix $\tilde{\mathbf C}$. Indeed,
if the smallest eigenvalue of $\tilde{\mathbf C}$ is larger than
$\Delta_0 / M$, then the constraint (\ref{Converted_to_real_1_3})
can not be satisfied for any estimate $\hat{\mathbf a} $. However,
$\Delta_0$ selected as suggested in Example~1 will satisfy the
feasibility condition that will guarantee the feasibility of the
problem
(\ref{Converted_to_real_1_1})--(\ref{Converted_to_real_1_3}).

\subsection{Semi-Definite Programming Relaxation}
If the problem
(\ref{Converted_to_real_1_1})--(\ref{Converted_to_real_1_3}) is
feasible, the equalities $\hat{\mathbf a} ^H \hat{\mathbf R}^{-1}
\hat{\mathbf a}  = Tr( \hat{\mathbf R}^{-1} \hat{\mathbf a}
\hat{\mathbf a} ^H) $ and $\hat{\mathbf a} ^H \tilde{\mathbf C}
\hat{\mathbf a}  = Tr( \tilde{\mathbf C} \hat{\mathbf a}
\hat{\mathbf a} ^H)$, where $Tr ( \cdot )$ denotes the trace of a
matrix, can be used to rewrite it as the following optimization
problem
\begin{eqnarray} \label{convnew1}
\min_{\hat{\mathbf a} } & & Tr ( \hat{\mathbf R}^{-1}
\hat{\mathbf a}  \hat{\mathbf a} ^H) \\
{\rm subject \ to} &&  Tr( \hat{\mathbf a}
\hat{\mathbf a} ^H) = M \\
&& Tr( \tilde{\mathbf C} \hat{\mathbf a}  \hat{\mathbf a} ^{H})
\le \Delta_0 . \label{convnew3}
\end{eqnarray}
Introducing the new variable $\mathbf A = \hat{\mathbf a}
\hat{\mathbf a}^H$, the problem
\eqref{convnew1}--\eqref{convnew3} can be casted as
\begin{eqnarray}
\min_{\mathbf A} && Tr ( \hat{\mathbf R}^{-1} \mathbf A)
\label{SDR_1_1} \\
{\rm subject \ to} && Tr(\mathbf A) = M \label{SDR_1_2} \\
&& Tr( \tilde{\mathbf C} \mathbf A) \le \Delta_0 \label{SDR_1_3} \\
&& rank(\mathbf A) = 1 \label{SDR_1_4}
\end{eqnarray}
where $rank ( \cdot )$ stands for the rank of a matrix and it is
guaranteed by the combination of the constraints \eqref{SDR_1_2}
and \eqref{SDR_1_4} that $\mathbf A$ is a positive semi-definite
matrix, i.e., $\mathbf A \succeq \mathbf 0$.

The only non-convex constraint in the problem
\eqref{SDR_1_1}--\eqref{SDR_1_4} is the rank-one constraint
(\ref{SDR_1_4}) while all other constraints and the objective are
convex. Using the SDP relaxation technique, the relaxed problem
can be obtained by dropping the non-convex rank-one constraint
(\ref{SDR_1_4}) and replacing it by the semi-definiteness
constraint $\mathbf A \succeq \mathbf 0$, which otherwise is not
guaranteed if (\ref{SDR_1_4}) is not present. Thus, the problem
\eqref{SDR_1_1}--\eqref{SDR_1_4} is replaced by the following
relaxed convex problem
\begin{eqnarray}
\min_{\mathbf A} && Tr ( \hat{\mathbf R}^{-1} \mathbf A)
\label{SDR_2_1} \\
{\rm subject \ to} && Tr( \mathbf A ) = M \label{SDR_2_2} \\
&& Tr ( \tilde{\mathbf C} \mathbf A) \le \Delta_0 \label{SDR_2_3} \\
&& \mathbf A \succeq \mathbf 0 . \label{SDR_2_4}
\end{eqnarray}

There are two features related to the use of SDR that have to be
addressed. First, it is possible, in general, that the original
problem is infeasible, however the relaxed one is feasible.
Second, the optimal solution of the relaxed problem is, in
general, an approximation of the optimal solution of the original
problem. Thus, it is desirable in general to estimate the
approximation bounds for the approximate solution and the
probability that both approximate and exact optimal solutions
coincide \cite{Tom}. In the following we will address these
issues.

\subsection{Feasibility and Rank of the Optimal Solution}
The result that connects the feasibility of the relaxed problem
(\ref{SDR_2_1})--(\ref{SDR_2_4}) to the feasibility of the
original problem
(\ref{Converted_to_real_1_1})-(\ref{Converted_to_real_1_3}) is
given in terms of the following theorem.

\textbf{Theorem 1:} {\it The problem
\eqref{SDR_2_1}--\eqref{SDR_2_4} is feasible if and only if the
problem
\eqref{Converted_to_real_1_1}--\eqref{Converted_to_real_1_3} is
feasible.}

\textbf{Proof:} See Appendix.

If the relaxed problem (\ref{SDR_2_1})--(\ref{SDR_2_4}) has a
rank-one solution, then the principle eigenvector of the solution
of \eqref{SDR_2_1}--\eqref{SDR_2_4}  will be the exact solution to
the original problem
(\ref{Converted_to_real_1_1})--(\ref{Converted_to_real_1_3}).
Otherwise, randomization procedures \cite{PhanVor}, \cite{Zhang}
have to be used, which can find the exact optimal solution of the
original problem only with a certain probability \cite{Tom}.
However, under the condition that the original optimization
problem
\eqref{Converted_to_real_1_1}--\eqref{Converted_to_real_1_3} or,
equivalently, the relaxed problem \eqref{SDR_2_1}--\eqref{SDR_2_4}
is feasible, the solution of the original problem can be extracted
from the solution of the relaxed problem by means of the following
constructive theorem.

\textbf{Theorem 2:} {\it Let $\mathbf A^*$ be the rank $r$ optimal
minimizer of the relaxed problem \eqref{SDR_2_1}--\eqref{SDR_2_4},
i.e., $\mathbf A^* = \mathbf Y \mathbf Y^{H}$ (where $\mathbf{Y}$
is an ${N \times r}$ matrix). If $r = 1$, the optimal solution of
the original problem simply equals $\mathbf Y$. Otherwise, it
equals $\mathbf Y \mathbf v$, where $\mathbf v$ is an $r \times 1$
vector such that $\| \mathbf Y \mathbf v \| = \sqrt{M}$ and
$\mathbf v^H \mathbf Y^H \tilde{\mathbf C} \mathbf Y \mathbf v =
Tr ( \mathbf Y^{H} \tilde{\mathbf C} \mathbf Y)$. Then one
possible solution for the vector $\mathbf v$ is proportional to
the sum of the eigenvectors of the following $r \times r$ matrix
\begin{equation} \label{OptSol2}
\mathbf D = \frac{1}{M} \mathbf Y^H \mathbf Y - \frac{\mathbf
Y^{H} \tilde{\mathbf C} \mathbf Y}{Tr ( \mathbf Y^{H}
\tilde{\mathbf C} \mathbf Y)}.
\end{equation} }

\textbf{Proof:} See Appendix.

Finally, we can prove the following result on the uniqueness of
the rank-one solution of the relaxed problem
(\ref{SDR_2_1})--(\ref{SDR_2_4}).

\textbf {Theorem 3:} { \it Under the condition that the solution
of the original optimization problem
\eqref{Converted_to_real_1_1}--\eqref{Converted_to_real_1_3} is
unique regardless of a phase shift, where the latter means that if
$\hat{\mathbf a} $ and $\hat{\mathbf a} ^\prime$ are both optimal
solutions, then there exists such phase shift $\phi$ that
$\hat{\mathbf a}  = \hat{\mathbf a} ^\prime e^{j \phi}$, the
solution of the relaxed problem (\ref{SDR_2_1})--(\ref{SDR_2_4})
always has rank one.}

\textbf{Proof:} See Appendix.

Note that the phase shift plays no role in the desired signal
steering vector estimation problem since the output power
\eqref{outPower} as well as the output SINR do not change if
$\hat{\mathbf a}$ undergo any phase rotation. Thus, the uniqueness
condition regardless a phase shift in Theorem~3 is proper . Under
this condition, the solution of the relaxed problem
(\ref{SDR_2_1})--(\ref{SDR_2_4}) is rank-one and the solution of
the original problem
\eqref{Converted_to_real_1_1}--\eqref{Converted_to_real_1_3} can
be found as a dominant eigenvector of the optimal solution of the
relaxed problem (\ref{SDR_2_1})--(\ref{SDR_2_4}). However, even
such uniqueness condition regardless a phase shift is not
necessarily satisfied for the problem
\eqref{Converted_to_real_1_1}--\eqref{Converted_to_real_1_3} (see
Example~2 below), and then we resort to the constructive
Theorem~2, which shows how to find the rank-one solution of
\eqref{Converted_to_real_1_1}--\eqref{Converted_to_real_1_3}
algebraically without any use of randomization procedures.

{\bf Example~2:} As an example of the situation when
\eqref{Converted_to_real_1_1}--\eqref{Converted_to_real_1_3} does
not have a unique solution, let us consider a ULA with 10
omni-directional antenna elements. The presumed direction of
arrival of the desired user is assumed to be $\theta_p=3^\circ$
with no interfering sources and the range of the desired signal
angular locations is equal to $ \Theta=[\theta_p - 12
^\circ,\theta_p + 12^\circ]$. The actual steering vector of the
desired user is perturbed due to the incoherent local scattering
effect and it can be expressed as $\mathbf a = \mathbf p + \mathbf
b$, where $\mathbf p = \mathbf d (3^\circ)$ is the steering vector
of the direct path and $\mathbf b$ is the steering vector of the
coherently scattered path. Let us consider the case when $\mathbf
b$ is orthogonal to $\mathbf p$. This later condition can be
satisfied if $\mathbf b$ is selected as $\mathbf d (-8.5 ^\circ)$.
In this case, both of the vectors $\mathbf p$ and $\mathbf b$ are
the eigenvectors of the matrix $\mathbf R ^{-1}$ which correspond
to the smallest eigenvalue. Since, these vectors satisfy the
constraints
(\ref{Converted_to_real_1_2})--(\ref{Converted_to_real_1_3}) and
correspond to the minimum eigenvalue, both of them are optimal
solutions of the optimization problem
(\ref{Converted_to_real_1_1})-(\ref{Converted_to_real_1_3}), thus,
the solution of the problem
\eqref{Converted_to_real_1_1}--\eqref{Converted_to_real_1_3} is
not unique.

\subsection{Solution Based on Strong Duality}
The solution of
\eqref{Converted_to_real_1_1}--\eqref{Converted_to_real_1_3} can
also be found using the strong duality theory. It follows from
Theorem~2 that the optimal value of the relaxed problem
(\ref{SDR_2_1})--(\ref{SDR_2_4}) is the same as the optimal value
of the the original problem
(\ref{Converted_to_real_1_1})--(\ref{Converted_to_real_1_3}). It
is because in addition to the fact that $\mathbf Y \mathbf v$ is
the optimal solution of the original problem
(\ref{Converted_to_real_1_1})--(\ref{Converted_to_real_1_3}) (see
Theorem~2),  $\mathbf Y \mathbf v (\mathbf Y \mathbf v)^H$ is also
the optimal solution of the relaxed problem
(\ref{SDR_2_1})--(\ref{SDR_2_4}) (see the proof of Theorem~2).
Furthermore, the dual problem of the the relaxed problem
(\ref{SDR_2_1})--(\ref{SDR_2_4}) is the same as the dual problem
of the original problem
(\ref{Converted_to_real_1_1})--(\ref{Converted_to_real_1_3}).
Indeed, by maximizing the dual function of the problem
(\ref{SDR_2_1})--(\ref{SDR_2_4}), which is the same as the dual
function of
(\ref{Converted_to_real_1_1})--(\ref{Converted_to_real_1_3}), the
dual problem for both the relaxed and original problems can be
written as
\begin{eqnarray}
\max_{\gamma_1, \gamma_2} && \gamma_1 M - \gamma_2 \Delta_0
\label{D_P_1} \\
{\rm subject \ to} && \hat{\mathbf R}^{-1} - \gamma_1 \mathbf I +
\gamma_2 \tilde{\mathbf C} \succeq \mathbf 0 \label{D_P_2}
\end{eqnarray}
where $\gamma_1$ and $\gamma_2 \geq 0 $ are the Lagrange
multipliers associated with constraints
(\ref{Converted_to_real_1_2}) and (\ref{Converted_to_real_1_3}) of
the original problem or the constraints (\ref{SDR_2_2}) and
(\ref{SDR_2_3}) of the relaxed problem, respectively. Since the
relaxed problem (\ref{SDR_2_1})--(\ref{SDR_2_4}) is convex, the
strong duality between (\ref{SDR_2_1})--(\ref{SDR_2_4}) and
(\ref{D_P_1})--(\ref{D_P_2}) holds, i.e., the optimal value of
(\ref{SDR_2_1})--(\ref{SDR_2_4}) is the same as the optimal value
of (\ref{D_P_1})--(\ref{D_P_2}). It implies that the optimal value
of the dual problem (\ref{D_P_1})--(\ref{D_P_2}) is also the same
as the optimal value of the original problem
(\ref{Converted_to_real_1_1})--(\ref{Converted_to_real_1_3}).
Thus, the strong duality between the dual problem
(\ref{D_P_1})--(\ref{D_P_2}) and the original problem
(\ref{Converted_to_real_1_1})--(\ref{Converted_to_real_1_3}) also
holds. It is worth mentioning that the strong duality of a
non-convex quadratic optimization problem with two positive
semi-definite quadratic constraints has been also studied in the
recent work \cite{beck}. Particularly, it has been shown that if a
non-convex quadratic optimization problem with two quadratic
constraints is strictly feasible, then strong duality holds. This
result agrees with our above conclusion that the strong duality
between (\ref{D_P_1})--(\ref{D_P_2}) and
(\ref{Converted_to_real_1_1})--(\ref{Converted_to_real_1_3})
holds. Indeed, it can be easily shown that if $\Delta_0 / M$ is
greater than the smallest eigenvalue of the matrix $\tilde{\mathbf
C}$, then the problem
(\ref{Converted_to_real_1_1})--(\ref{Converted_to_real_1_3}) is
strictly feasible, and thus the result of \cite{beck},
\cite{beck2} applies and the strong duality between
(\ref{D_P_1})--(\ref{D_P_2}) and
(\ref{Converted_to_real_1_1})--(\ref{Converted_to_real_1_3})
holds. Moreover, if $\Delta_0 / M$ is equal to the smallest
eigenvalue of the matrix $\tilde{\mathbf C}$, the problem
(\ref{Converted_to_real_1_1})--(\ref{Converted_to_real_1_3}) has
limited number of feasible points and it can be solved easily by
checking all these points. Thus, it is simply assumed in the
sequel that $\Delta_0 / M$ is greater than the smallest eigenvalue
of $\tilde{\mathbf C}$.

The dual problem (\ref{D_P_1})--(\ref{D_P_2}) belongs to the class
of SDP problems and, thus, can be solved efficiently using, for
example, interior-point methods. Moreover, it contains only two
optimization variables. Let the optimal solution of the the dual
problem (\ref{D_P_1})--(\ref{D_P_2}) be $\gamma_1^*$ and
$\gamma_2^*$. It is easy to see that $\gamma_1^*$ is always
strictly positive and the matrix $\hat{\mathbf R}^{-1} -
\gamma_1^* \mathbf I + \gamma_2^* \tilde{\mathbf C}$ is rank
deficient. Indeed, in order to maximize the objective function
\eqref{D_P_1} for a fixed $\gamma_2 \geq 0$, $\gamma_1$ should be
equal to the smallest eigenvalue of the matrix $\hat{\mathbf
R}^{-1} + \gamma_2 \tilde{\mathbf C}$ which makes the matrix $
\hat{\mathbf R}^{-1} - \gamma_1 \mathbf I + \gamma_2
\tilde{\mathbf C}$ rank deficient. Furthermore, since for every
nonnegative $\gamma_2$, $\hat{\mathbf R}^{-1} + \gamma_2
\tilde{\mathbf C}$ is a positive definite matrix, we obtain that
$\gamma_1^*$ is positive and $ \hat{\mathbf R}^{-1} - \gamma_1
\mathbf I + \gamma_2 \tilde{\mathbf C}$ is rank deficient. Since
the strong duality between
(\ref{Converted_to_real_1_1})--(\ref{Converted_to_real_1_3}) and
(\ref{D_P_1})--(\ref{D_P_2}) holds, the necessary and sufficient
optimality conditions can be written as

\begin{eqnarray}
&& ( \hat{\mathbf R}^{-1} - \gamma_1^* \mathbf I + \gamma_2^*
\tilde{\mathbf C}) \hat{\mathbf a}  = \mathbf{0}_{M }
\label{NSC_2_1} \\
&& \hat{\mathbf a} ^H \hat{\mathbf a}  = M
\label{NSC_2_2} \\
&& \gamma_2^* (\hat{\mathbf a} ^H \tilde{\mathbf C}
\hat{\mathbf a}  - \Delta_0 ) = 0 \label{NSC_2_4} \\
&& \hat{\mathbf a} ^H \tilde{\mathbf C} \hat{\mathbf a} \leq
\Delta_0 . \label{NSC_2_3}
\end{eqnarray}
where $\mathbf{0}_{M }$ is the vector of zeros of length $M$.
Moreover, using the fact that $\gamma_1^*$ is strictly positive
and the matrix $\hat{\mathbf R}^{-1} - \gamma_1^* \mathbf I +
\gamma_2^* \tilde{\mathbf C}$ is rank deficient, the solution of
the original problem
(\ref{Converted_to_real_1_1})--(\ref{Converted_to_real_1_3}) can
easily be found.\footnote{Note that the general form of the
optimality conditions \eqref{NSC_2_1}-\eqref{NSC_2_3} has been
solved in \cite{beck}.} There are two possible situations.
\begin{itemize}
\item[(i)] The matrix $\hat{\mathbf R}^{-1} -\gamma_1^* \mathbf I +
\gamma_2^* \tilde{\mathbf C} $ has only one zero eigenvalue. In
this case, the only vector which satisfies \eqref{NSC_2_1} is
given by
\begin{equation}
\hat{\mathbf a}  = \sqrt{M} \boldsymbol{\rho} \left\{ \hat{\mathbf
R}^{-1} - \gamma_1^* \mathbf I + \gamma_2^* \tilde{\mathbf C}
\right\} \label{optsol}
\end{equation}
where $\boldsymbol{\rho} \{\cdot\}$ denotes the eigenvector of a
matrix which corresponds to the smallest eigenvalue. Indeed,
\eqref{optsol} satisfies the necessary and sufficient optimality
conditions \eqref{NSC_2_1}--\eqref{NSC_2_3} and, thus, is the
optimal solution of
(\ref{Converted_to_real_1_1})--(\ref{Converted_to_real_1_3}).

\item[(ii)]  The matrix $\hat{\mathbf R}^{-1} -\gamma_1^* \mathbf I +
\gamma_2^* \tilde{\mathbf C} $ has more than one zero eigenvalue.
In this case, consider the matrix $\mathbf F$ each column of which
is an eigenvector of the matrix $\hat{\mathbf R}^{-1} - \gamma_1^*
\mathbf I + \gamma_2^* \tilde{\mathbf C} $ corresponding to zero
eigenvalue. Thus, the dimension of $\mathbf F$ is $M \times q$
where $q$ is the number of zero eigenvalues of the matrix
$\hat{\mathbf R}^{-1} -\gamma_1^* \mathbf I + \gamma_2^*
\tilde{\mathbf C}$. Then vectors $\hat{\mathbf a} $ that satisfy
the condition (\ref{NSC_2_1}) can be written as
\begin{equation}
\hat{\mathbf a}  = \mathbf F \mathbf q
\end{equation}
where $\mathbf q$ is a ${k \times 1}$ vector. The optimality
conditions \eqref{NSC_2_1}--\eqref{NSC_2_3} can be rewritten then
in terms of $\mathbf q$ and $\mathbf F$ as
\begin{eqnarray}
&& \mathbf q^H \mathbf q = M \label{NSC_3_1} \\
&& \gamma_2^* (\mathbf q^H \mathbf F^H \tilde{\mathbf C} \mathbf F
\mathbf q - \Delta_0 ) = 0 \label{NSC_3_3} \\
&& \mathbf q^H \mathbf F^H \tilde{\mathbf C} \mathbf F \mathbf q
\leq \Delta_0 \label{NSC_3_2} .
\end{eqnarray}
Let $\mu_{max}$ and $\mu_{min}$ denote the largest and the
smallest eigenvalues of the matrix $\mathbf F^H \tilde{\mathbf C}
\mathbf F$ and $\mathbf f_{max}$ and $\mathbf f_{min}$ stand for
their corresponding eigenvectors. Then the following two subcases
should be considered.
\begin{itemize}
\item[(ii.a)] The first subcase is when $\gamma_2^* = 0$. Then
$\mu_{min} \leq \Delta_0 / M$ and $\mathbf q$ can be simply chosen as
\begin{equation}
\mathbf q = \sqrt{M} \mathbf f_{min} .
\end{equation}

\item[(ii.b)] The other subcase is when $\gamma_2^* > 0$.
Then \eqref{NSC_3_3} implies that $\mathbf q^H \mathbf F^H
\tilde{\mathbf C} \mathbf F \mathbf q = \Delta_0$ and as in the
previous subcase $\mu_{min} \leq \Delta_0 / M$. Moreover,
$\mu_{max} \geq \Delta_0 / M$. Therefore, $\mathbf q$ can be
chosen as a linear combination of $\mathbf f_{min}$ and $\mathbf
f_{max}$ as follows
\begin{equation}
\mathbf q = \sqrt{M} ( \sqrt{1 - \theta} \mathbf f_{min} +
\sqrt{\theta} \mathbf f_{max})
\end{equation}
where $\theta = ( \Delta_0 / M - \mu_{min} ) / ( \mu_{max} -
\mu_{min} )$.
\end{itemize}
\end{itemize}

As soon as the estimate $\hat{\mathbf a} $ is obtained, the
beamforming weight vector can be straightforwardly computed as
\begin{equation} \label{BemafSDR}
\mathbf w_{\rm SDP} = \alpha^\prime \hat{\mathbf R}^{-1}
\hat{\mathbf a}
\end{equation}
where $\alpha^\prime = 1 / \hat{\mathbf a} ^H \hat{\mathbf R}^{-1}
\hat{\mathbf a} $. The beamformer \eqref{BemafSDR} can be compared
to the eigenspace-based robust adaptive beamforming
\eqref{eigenbeam} where the imprecisely known signal steering
vector is corrected by projecting it to the
signal-plus-interference subspace. However, a significant
difference is that no knowledge of the dimension of the
signal-plus-interference subspace is needed in the proposed
beamforming as well as no subspace swap can happen at low SNRs as
in \eqref{eigenbeam}.

\section{Special Cases and Relationships}

\subsection{Simple Solution Under the Constraint \eqref{NasrConst}}
For high and moderate SNRs, the protection against convergence to
an interference steering vector can be ensured by means of the
constraint \eqref{NasrConst} (see also \cite{Vorobyov5}). Then the
corresponding desired signal steering vector estimation problem
can be written as
\begin{eqnarray}
\min_{ \hat{\mathbf a}  } && \hat{\mathbf a} ^H \hat{\mathbf
R}^{-1} \hat{\mathbf a}
\label{eq:estimation1} \\
{\rm subject \ to} && \left\| \hat{\mathbf a}  \right\| =
\sqrt{M} \label{eq:estimation2} \\
&& \mathbf P^\bot \hat{\mathbf a}  = 0 . \label{eq:estimation3}
\end{eqnarray}
This problem differs from the problem
(\ref{Converted_to_real_1_1})--(\ref {Converted_to_real_1_3}) only by
the constraint \eqref{eq:estimation3}. The problem
(\ref{eq:estimation1})--(\ref{eq:estimation3}) is a non-convex
optimization problem, but a simple closed-form solution can be
found. The main idea is to first find a set of vectors satisfying
the constraint (\ref{eq:estimation3}). Note that $\mathbf P^\bot
\hat{\mathbf a} = 0$ implies that $\hat{\mathbf a} = \mathbf U
\mathbf U^H \hat{\mathbf a} $ and, therefore, we can write that
\begin{eqnarray}
\hat{\mathbf a}  = \mathbf U \mathbf v \label{eq:solset1}
\end{eqnarray}
where $\mathbf v$ is a $L \times 1$ complex valued vector. Using
(\ref{eq:solset1}), the optimization problem
(\ref{eq:estimation1})--(\ref{eq:estimation3}) can be equivalently
rewritten in terms of $\mathbf v$ as
\begin{eqnarray}
\min_\mathbf v && \mathbf v^H \mathbf U^H \hat{\mathbf R}^{-1}
\mathbf U \mathbf v \label{eq:equ1} \\
{\rm subject \ to} &&  \| \mathbf v \| = \sqrt{M} .
\label{eq:equ2}
\end{eqnarray}
Finally, the solution to the optimization problem of type
(\ref{eq:equ1})--(\ref{eq:equ2}) is known to be the eigenvector of
the matrix $\mathbf U^H \hat{\mathbf R}^{-1}\mathbf U$ which
corresponds to the minimum eigenvalue. Thus, the estimate of the
steering vector can be obtained as
\begin{eqnarray}
\hat{\mathbf a}  = \mathbf U \cdot \boldsymbol{\rho} \left\{
\mathbf U^H \hat{\mathbf R}^{-1} \mathbf U \right\}
\label{eq:optsol}
\end{eqnarray}
and the corresponding beamforming vector is
\begin{eqnarray}
\mathbf w_{\rm 1} = \alpha^\prime \hat{\mathbf R}^{-1} \mathbf U
\cdot \boldsymbol{\rho} \left\{ \mathbf U^H \hat{\mathbf R}^{-1}
\mathbf U \right\} . \label{eq:optsolW}
\end{eqnarray}
As compared to the the eigenspace-based robust adaptive
beamforming \eqref{eigenbeam}, the steering vector in the
beamformer \eqref{eq:optsolW} is the eigenvector corresponding to
the smallest eigenvalue of the matrix $\mathbf U^H \hat{\mathbf
R}^{-1} \mathbf U$.

\subsection{Signal-to-Interference Ratio $\rightarrow \infty$}
In the case when the signal-to-interference ratio approaches
infinity (SIR$\rightarrow \infty$), it is guaranteed that the
estimate of the desired signal steering vector will not converge
to an interference steering vector and, thus, the constraint
(\ref{Converted_to_real_1_3}) is never active and can be dropped.
Then the optimization problem
(\ref{Converted_to_real_1_1})--(\ref{Converted_to_real_1_3})
simplifies as
\begin{eqnarray}
\min_{\hat{\mathbf a} } && \hat{\mathbf a} ^H
\hat{\mathbf R}^{-1} \hat{\mathbf a}  \label{high_snr_1_1} \\
{\rm subject \ to} && \left\| \hat{\mathbf a}  \right\| = \sqrt{M}
. \label{high_snr_1_2}
\end{eqnarray}
The solution of the latter problem is $\hat{\mathbf a} = \sqrt{M}
\boldsymbol \rho (\hat{\mathbf R}^{-1} )$. Interestingly, the
optimization problem \eqref{high_snr_1_1}--\eqref{high_snr_1_2} is
the same as the optimization problem \cite[(39)]{LiStoicaWang}
which is obtained after dropping the constraint $\| \boldsymbol
\delta \| \leq \varepsilon$. This relationship holds
only for SIR$\rightarrow \infty$ when no additional constraints
are required to guarantee that the estimate of the steering vector
does not converge to an interference steering vector. In this
case, the proposed and the the worst-case-based beamformer are the
same, that is,
\begin{eqnarray}
\mathbf w_{\rm 2} = \alpha^\prime \hat{\mathbf R}^{-1}
\hat{\mathbf a}  = \alpha^\prime \hat{\mathbf R}^{-1}
\boldsymbol{\rho} \left\{ \hat{\mathbf R}^{-1} \right\} .
\label{eq:optsolWa}
\end{eqnarray}

\section{Simulation Results}
Throughout the simulations, a ULA of $10$ omni-directional antenna
elements with the inter-element spacing of half wavelength is
considered. Additive noise in antenna elements is modeled as
spatially and temporally independent complex Gaussian noise with
zero mean and unit variance. Two interfering sources are assumed
to imping on the antenna array from the directions $30^\circ$ and
$50^\circ$, while the presumed direction towards the desired
signal is assumed to be $3^\circ$. In all simulation examples, the
interference-to-noise ratio (INR) equals $30$~dB and the desired
signal is always present in the training data. For obtaining each
point in the examples, $100$ independent runs are used.

The proposed SDP relaxation-based beamformer is compared with
three other methods in terms of the output SINR. These robust
adaptive beamformers are (i)~the worst-case-based robust adaptive
beamformer \eqref{WCBsol}, (ii)~the SQP-based beamformer of
\cite{Vorobyov5}, and (iii)~the eigenspace-based beamformer
\eqref{eigenbeam}. For the proposed beamformer and the SQP-based
beamformer of \cite{Vorobyov5}, the angular sector of interest
$\Theta$ is assumed to be $\Theta =[ \theta_p - 5^\circ,\theta_p +
5^\circ]$ where $\theta_p$ is the presumed direction of arrival of
the desired signal. The CVX MATLAB toolbox \cite{cvx} is used for
solving the optimization problems (\ref{SDR_2_1})--(\ref{SDR_2_4})
and (\ref{D_P_1})--(\ref{D_P_2}) and the value of $\Delta_0$ is
set equal to the maximum value of the $\mathbf d^H (\theta)
\tilde{\mathbf C} \mathbf d(\theta)$ within the angular sector of
interest $\Theta$. The value $\delta = 0.1 $ and $6$ dominant
eigenvectors of the matrix $\mathbf C$ are used in the SQP-based
beamformer and the value $\varepsilon = 0.3 M$ is used for the
worst-case-based beamformer as it has been recommended in
\cite{Vorobyov1}. The dimension of the signal-plus-interference
subspace is assumed to be always estimated correctly for the
eigenspace-based beamformer and equals 3.

\subsection{Example 1: Exactly known signal steering vector}
\begin{figure}[p]
\begin{center}
\includegraphics[scale=.3]{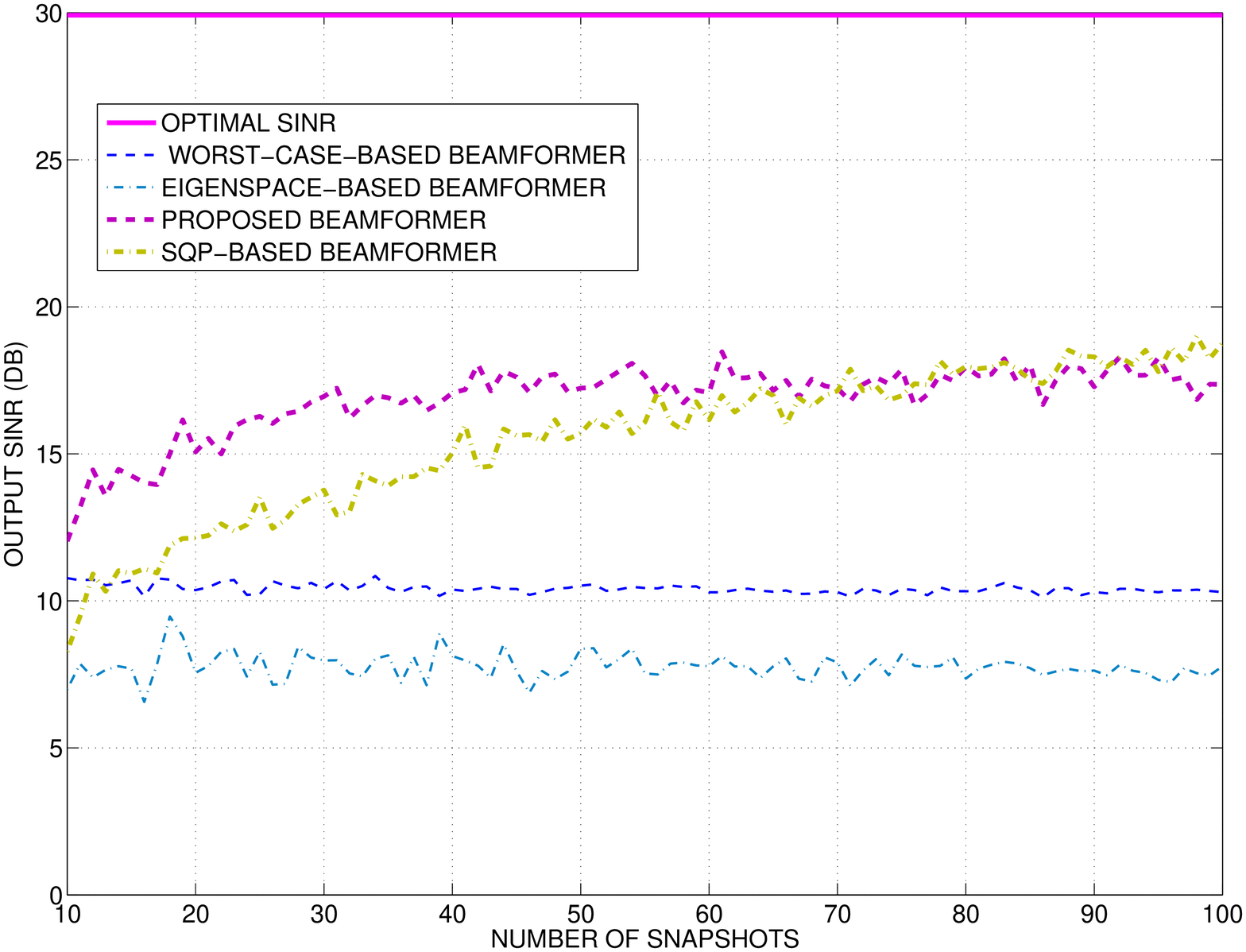}
\caption{Output SINR versus training sample size $K$ for fixed
$\rm{SNR}=20$~dB and $\rm{INR}=30$~dB .} \label{Fig1}
\end{center}
\end{figure}

\begin{figure}[p]
\begin{center}
\includegraphics[scale=.3]{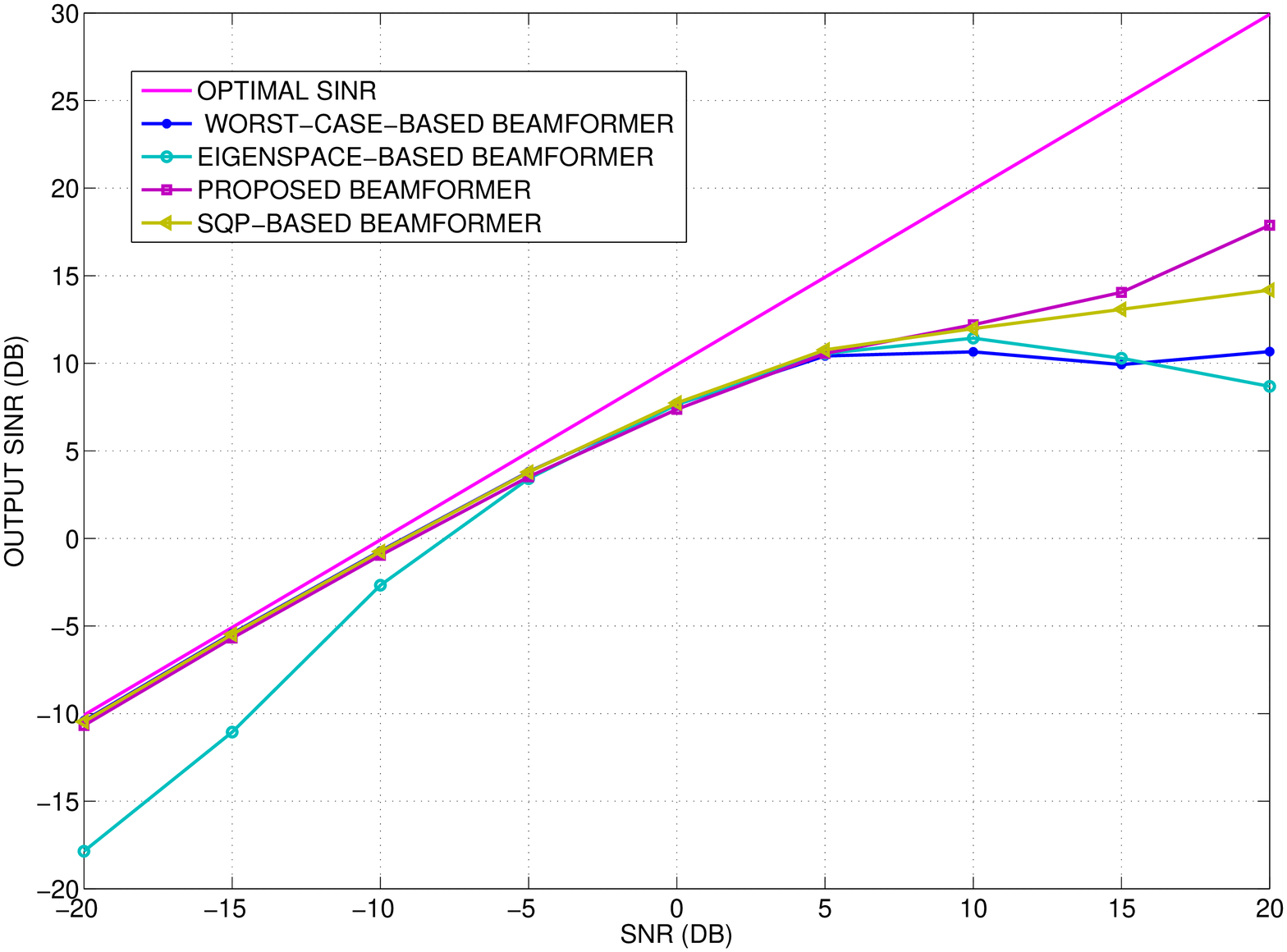}
\caption{Output SINR versus SNR for training data size of $K = 30$
and $\rm{INR}=30$~dB.} \label{Fig2}
\end{center}
\end{figure}

In the first example, we consider the case when the actual desired
signal steering vector is known exactly. Even in this case, the
presence of the signal of interest in the training data can
substantially reduce the convergence rates of adaptive beamforming
algorithms as compared to the signal-free training data case
\cite{Feldman}. In Fig.~\ref{Fig1}, the mean output SINRs for the
aforementioned methods are illustrated versus the number of
training snapshots for the fixed single-sensor $\rm{SNR}=20$~dB.
Fig.~\ref{Fig2} displays the mean output SINR of the same methods
versus the SNR for fixed training data size of $K = 30$. It can be
seen from these figures that the proposed beamforming technique
outperforms the other techniques even in the case of exactly known
signal steering vector. It is especially true for small sample
size.


\subsection{ Example 2: Signal Spatial Signature Mismatch Due to
Wavefront Distortion.} In the second example, we consider the
situation when the signal steering vector is distorted by wave
propagation effects in an inhomogeneous medium.
Independent-increment phase distortions are accumulated by the
components of the presumed steering vector. It is assumed that the
phase increments remain fixed in each simulation run and are
independently chosen from a Guassian random generator with zero mean
and variance $0.04$.

\begin{figure}[p]
\begin{center}
\includegraphics[scale=.3]{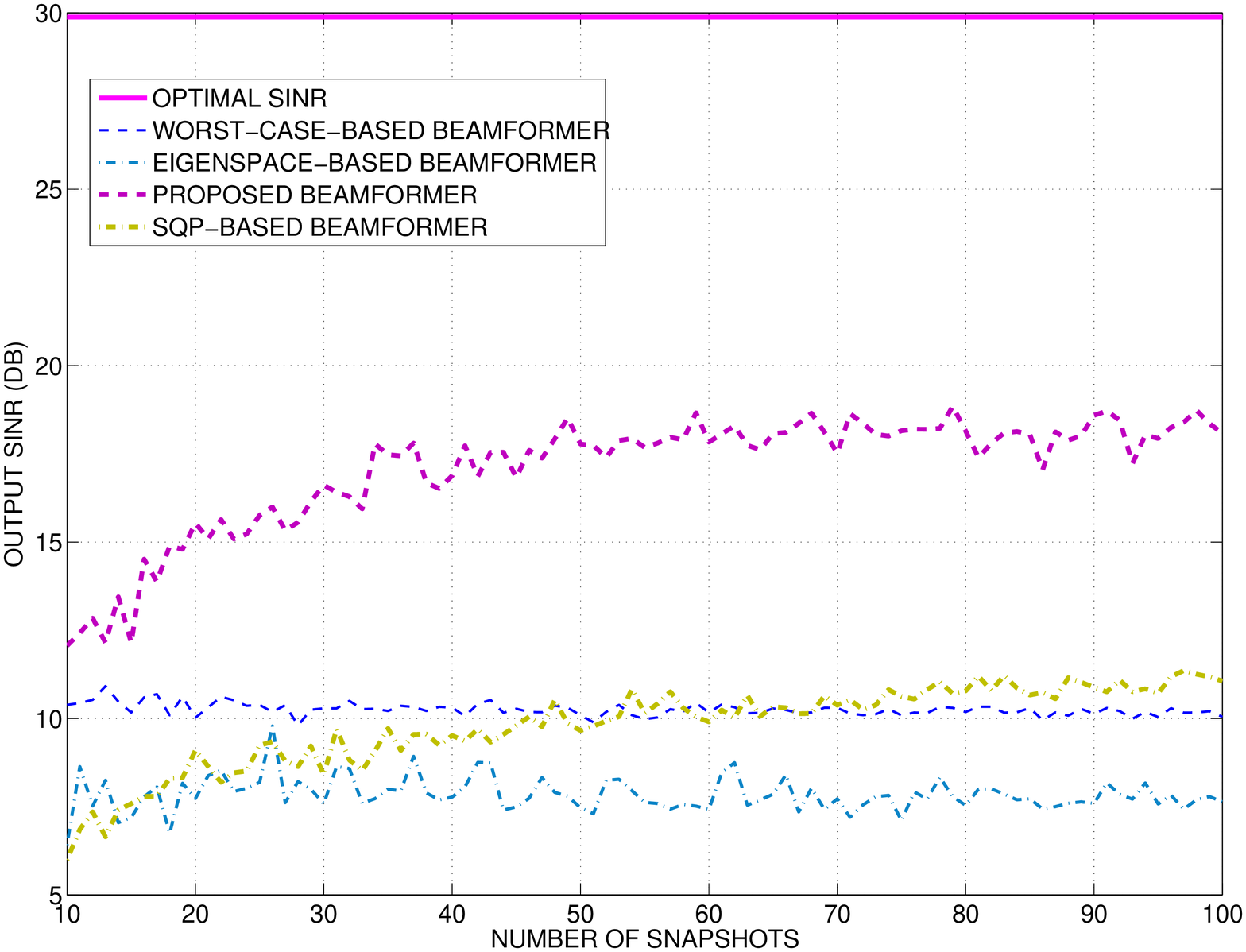}
\caption{Output SINR versus training sample size $K$ for fixed
$\rm{SNR}=20$~dB and $\rm{INR}=30$~dB.} \label{Fig3}
\end{center}
\end{figure}

\begin{figure}[p]
\begin{center}
\includegraphics[scale=.3]{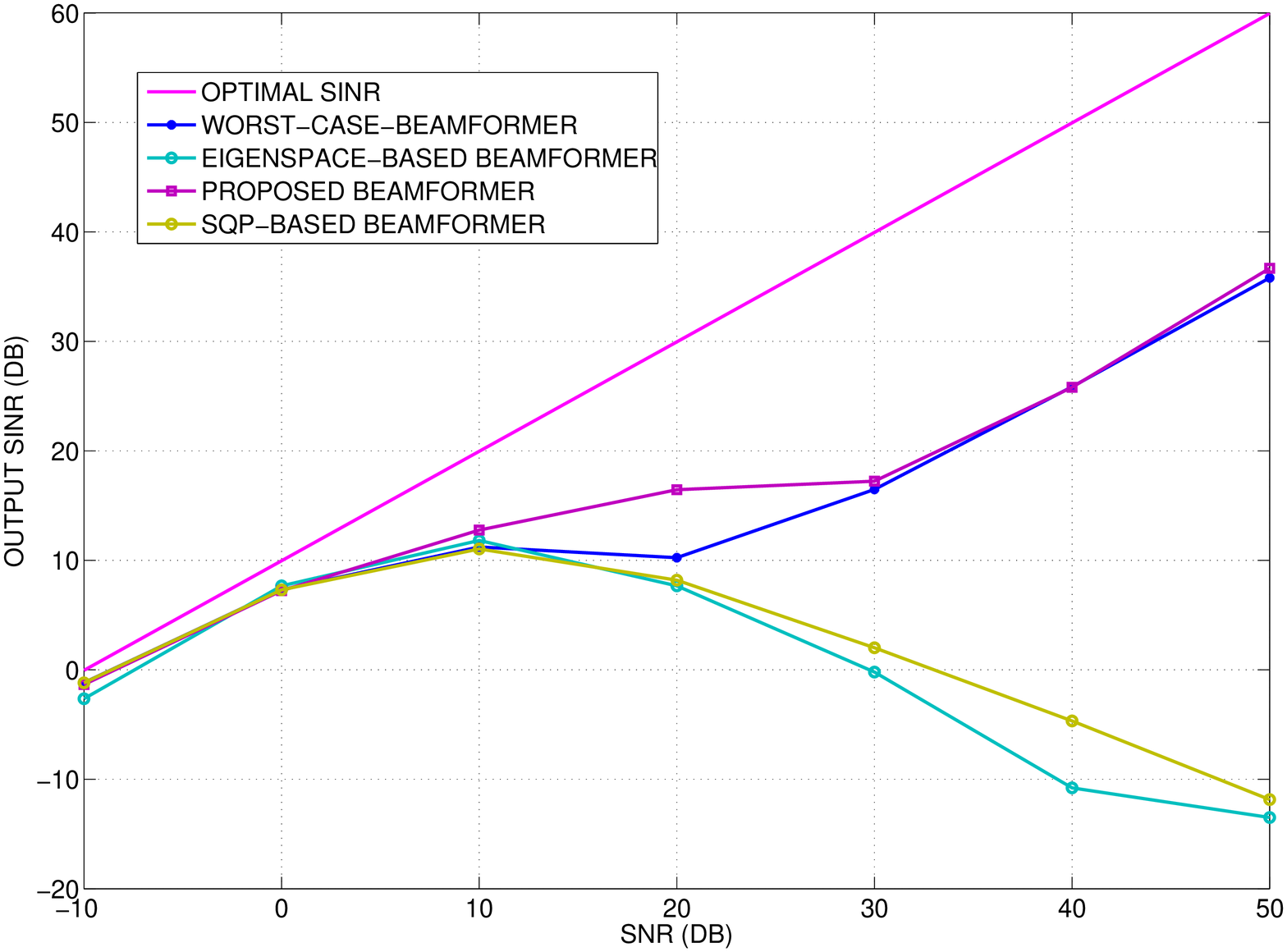}
\caption{Output SINR versus SNR for training data size of $K = 30$
and $\rm{INR}=30$~dB.} \label{Fignew}
\end{center}
\end{figure}

The performance of the methods tested is shown versus the number
of training snapshots for fixed single-sensor SNR$=20$~dB in
Fig.~\ref{Fig3} and versus the SNR for fixed training data size
$K=30$ in Fig.~\ref{Fignew}. It can be seen from these figures
that the proposed beamforming technique outperforms all other
beamforming techniques. Interestingly, it outperforms the
eigenspace-based beamformer even at high SNR. This performance
improvement compared to the eigenspace-based beamformer can be
attributed to the fact that the knowledge of sector which includes
the desired signal steering vector is used in the proposed
beamforming technique. Fig.~\ref{Fignew} also illustrates the case
when SNR$\gg$INR where INR stands for interference-to-noise ratio.
This case aims to illustrate the situation when SIR$\rightarrow
\infty$. As it can be expected, the proposed and the
worst-case-based methods perform almost equivalently.


\subsection{ Example 3: Signal Spatial Signature Mismatch Due
to Coherent Local Scattering} The third example corresponds to the
scenario of coherent local scattering \cite{coherentscattering}.
In this case, the desired signal steering vector is distorted by
local scattering effects so that the presumed signal
steering vector is a plane wave, whereas the actual steering
vector is formed by five signal paths as
\begin{equation}
\mathbf a = \mathbf p + \sum_{i=1}^{4} e^{j \psi_i} \mathbf b(
\theta_i)
\end{equation}
where $\mathbf p$ corresponds to the direct path and $\mathbf b(
\theta_i), \  i= 1,2,3, 4$ correspond to the coherently scattered
paths. We model the $i$th path $\mathbf b( \theta_i)$ as a a plane
wave impinging on the array from the direction $\theta_i$. The
angles $\theta_i$, $i=1,2,3,4$ are independently drawn in each
simulation run from a uniform random generator with mean $3^\circ$
and standard deviation $1^\circ$. The parameters $\psi_i,\ \  i=1,
2, 3, 4$ represent path phases that are independently and
uniformly drawn from the interval $[0, 2 \pi ]$ in each simulation
run. Note that $\theta_i$ and $\psi_i,\ \ i=1, 2, 3, 4$ change
from run to run but do not change from snapshot to snapshot.

\begin{figure}[p]
\begin{center}
\includegraphics[scale=.3]{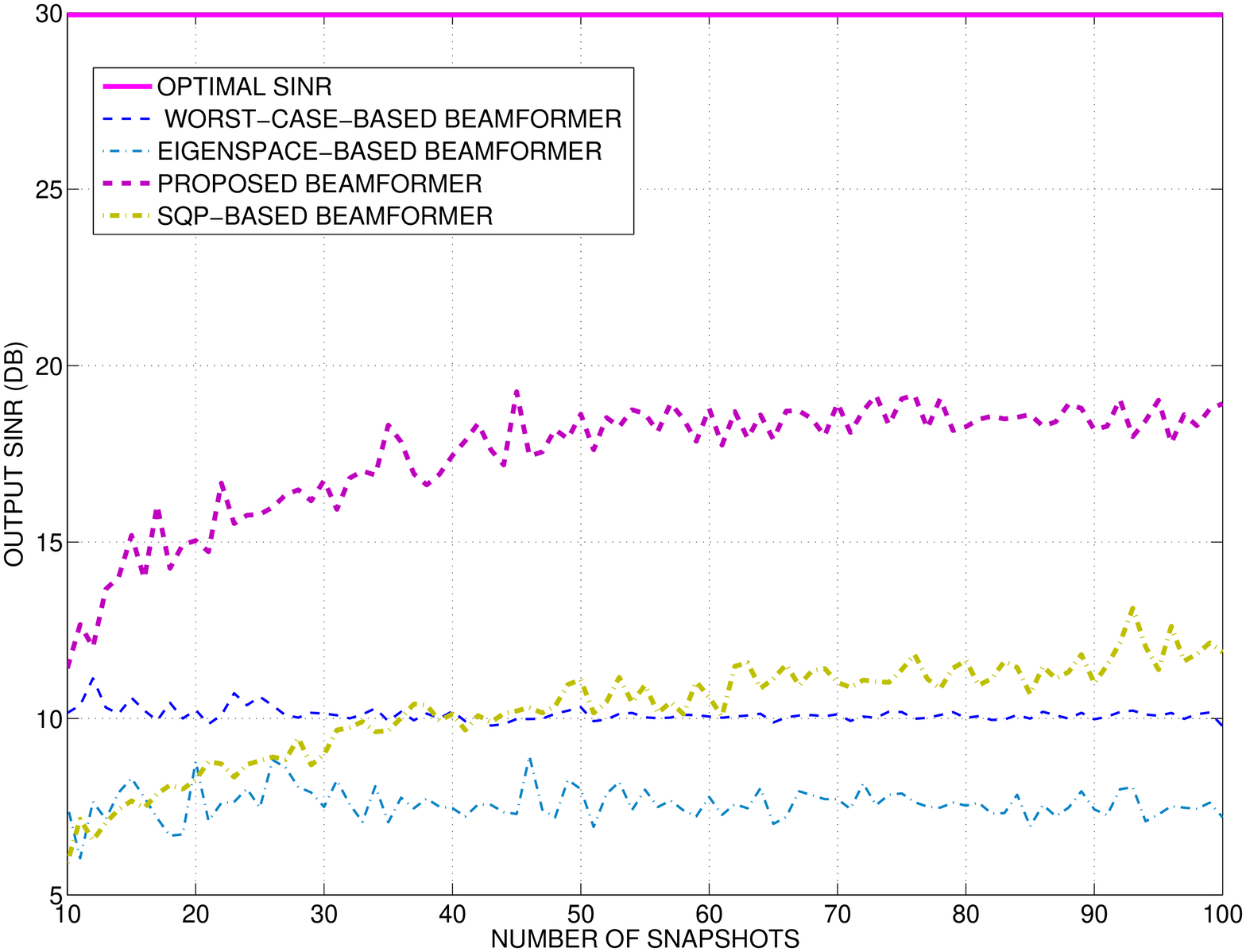}
\caption{Output SINR versus training sample size $K$ for fixed
$\rm{SNR}=20$~dB and $\rm{INR}=30$~dB.} \label{Fig5}
\end{center}
\end{figure}

\begin{figure}[p]
\begin{center}
\includegraphics[scale=.3]{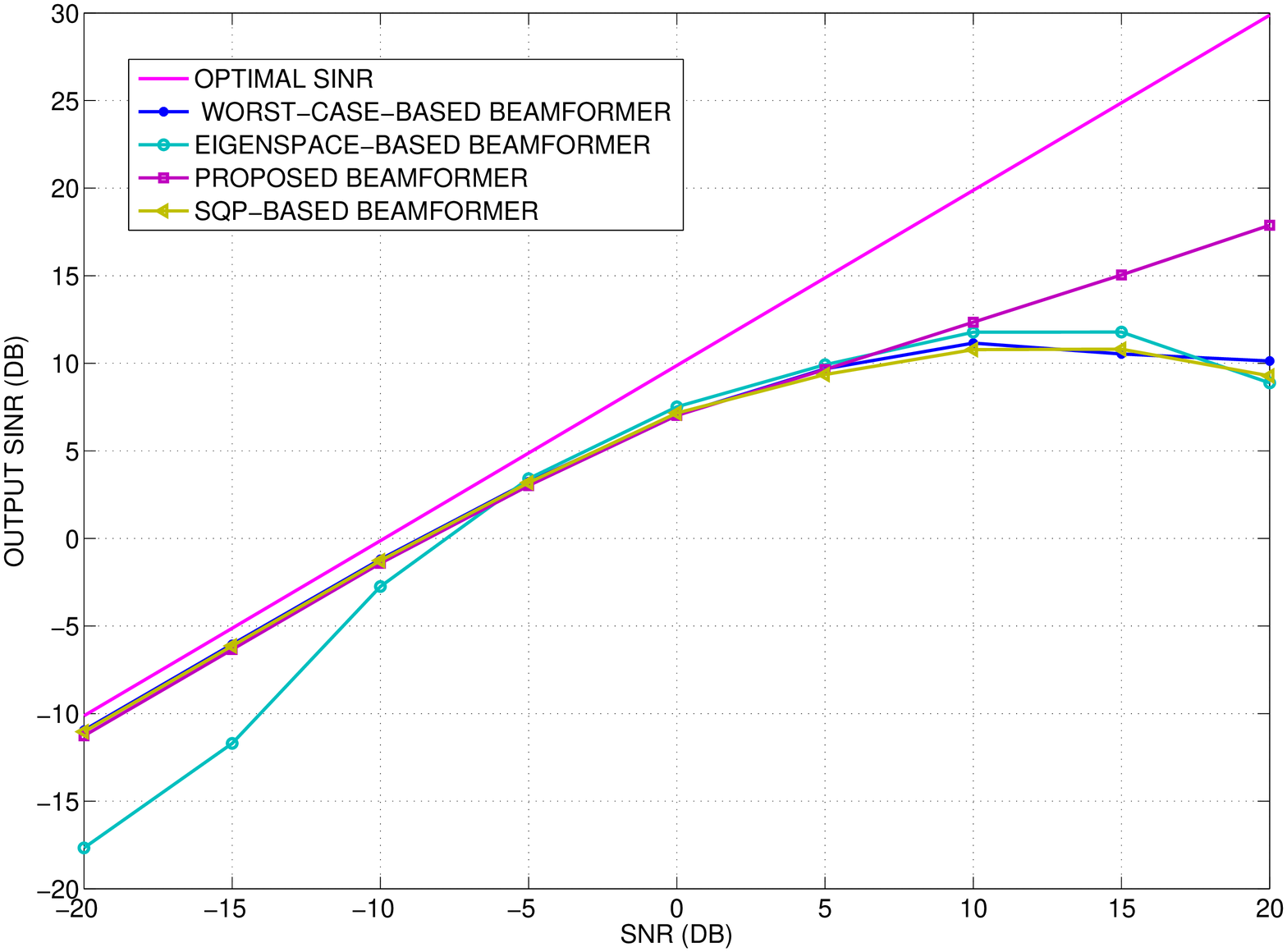}
\caption{Output SINR versus SNR  for training data size of $K =
30$ and $\rm{INR}=30$~dB.} \label{Fig6}
\end{center}
\end{figure}

Fig.~\ref{Fig5} displays the performance of all four methods
tested versus the number of training snapshots $K$ for fixed
single-sensor $\rm{SNR}= 20$~dB. Note that the SNR in this example
is defined by taking into account all signal paths. The
performance of the same methods versus SNR for the fixed
training data size $K=30 $ is displayed in Fig.~\ref{Fig6}.
Similar to the previous example, the proposed beamformer
significantly outperforms other beamformers due to its ability to
estimate the actual steering vector with a hight accuracy.


\section{Conclusion}
A new approach to robust adaptive beamforming in the presence of
signal steering vector errors has been developed. According to
this approach, the actual steering vector is first estimated using
its presumed (prior) value, and then this estimate is used to find
the optimal beamformer weight vector. The problem of signal
steering vector estimation belongs to the class of homogeneous
QCQP problems. It has been shown that this problem can be solved
using the SDP relaxation technique or the strong duality theory and the exact
solution for the signal steering vector can be found efficiently
or even in some cases in closed-form. As compared to another
well-known robust adaptive beamforming method based on the signal
steering vector correction/estimation, that is, the
eigespace-based method, the proposed technique does not suffer
from the subspace swap phenomenon since it does not use eigenvalue
decomposition of the sample covariance matrix. Moreover, it does
not require any knowledge on the number of interferences. As
compared to the well-known worst-case-based and
probabilistically-constrained robust adaptive beamformers, the
proposed technique does not use any design parameters of non-unique choice.
Our simulation results demonstrate the superior performance for
the proposed method over the aforementioned robust adaptive
beamforming methods for several frequently encountered types of
signal steering vector errors.

\section*{Appendix}
\subsection*{Proof of Theorem 1}
Let $\hat{\mathbf a} $ be a feasible point for the original
problem
(\ref{Converted_to_real_1_1})--(\ref{Converted_to_real_1_3}). It
is straightforward to see that $\mathbf A = \hat{\mathbf a}
\hat{\mathbf a} ^H$ is also a feasible point for the relaxed
problem (\ref{SDR_2_1})--(\ref{SDR_2_4}). Thus, the necessity
statement of the theorem follows trivially.

Let us now prove sufficiency. Let $\mathbf A = \sum_{i=1}^{M}
\lambda_i \mathbf b_i \mathbf b_i^H$ be a feasible point for the
relaxed problem (\ref{SDR_2_1})--(\ref{SDR_2_4}), where
$\lambda_i$ and $\mathbf b_i$, $i = 1, \cdots, M$ are,
respectively, the eigenvalues and eigenvectors of $\mathbf A$.
Also let $l = \arg \min_{i} {\mathbf b_i^H \tilde{\mathbf C}
\mathbf b_i}$ and $\hat{\mathbf a}  = \sqrt{M} \mathbf b_l$. Then
the following holds
\begin{eqnarray}
\hat{\mathbf a} ^H \hat{\mathbf a}  = M \mathbf b_l^H \mathbf b_l
= M
\end{eqnarray}
and the constraint (\ref{Converted_to_real_1_2}) is satisfied.

Moreover, we can write that
\begin{eqnarray} \label{eq1}
\hat{\mathbf a} ^H \tilde{\mathbf C} \hat{\mathbf a}  = M \mathbf
b_l^H \tilde{\mathbf C} \mathbf b_l = \left( \sum_{i=1}^{M}
\lambda_{i} \right) \mathbf b_l^H \tilde{\mathbf C} \mathbf b_l
\end{eqnarray}
where $\sum_{i=1}^{M} \lambda_{i} = Tr ( \mathbf A ) = M$. Using
the following inequality
\begin{eqnarray}
\left( \sum_{i=1}^{M} \lambda_{i} \right) \mathbf b_l^H
\tilde{\mathbf C} \mathbf b_l \leq \sum_{i=1}^{M} \lambda_{i}
\mathbf b_i^H \tilde{\mathbf C} \mathbf b_i
\end{eqnarray}
and \eqref{eq1}, we obtain that
\begin{eqnarray} \label{eq3}
\hat{\mathbf a} ^H \tilde{\mathbf C} \hat{\mathbf a}  \leq
\sum_{i=1}^{M} \lambda_{i} \mathbf b_i^H \tilde{\mathbf C} \mathbf
b_i .
\end{eqnarray}
The right hand side of \eqref{eq3} can be further rewritten as
\begin{eqnarray} \label{eq4}
\sum_{i=1}^{M} \lambda_{i} \mathbf b_i^H \tilde{\mathbf C} \mathbf
b_i = \sum_{i=1}^{M} \lambda_{i} Tr( \tilde{\mathbf C} \mathbf b_i
\mathbf b_i^H ).
\end{eqnarray}
Using the property of the trace that a sum of traces is equal to
the trace of a sum, we obtain that
\begin{eqnarray} \label{eq5}
\sum_{i=1}^{M} \lambda_{i} Tr( \tilde{\mathbf C} \mathbf b_i
\mathbf b_i^H ) = Tr \left( \tilde{\mathbf C} \sum_{i=1}^{K}   (
\lambda_{i}  \mathbf b_i \mathbf b_i^H) \right) .
\end{eqnarray}
Moreover, since $\sum_{i=1}^{K} \lambda_{i} \mathbf b_i \mathbf
b_i^H = \mathbf A$, we have
\begin{eqnarray} \label{eq6}
\sum_{i=1}^{M} \lambda_{i} \mathbf b_i^H \tilde{\mathbf C} \mathbf
b_i = Tr \left( \tilde{\mathbf C} \sum_{i=1}^{K} \lambda_{i}
\mathbf b_i \mathbf b_i^H \right) = Tr (\tilde{\mathbf C} \mathbf
A) .
\end{eqnarray}
Substituting \eqref{eq6} in the left hand side of \eqref{eq3}, we
finally obtain that
\begin{eqnarray} \label{eq7}
\hat{\mathbf a} ^H \tilde{\mathbf C} \hat{\mathbf a}  \leq Tr
(\tilde{\mathbf C} \mathbf A) \leq \Delta_0 .
\end{eqnarray}
Therefore, the constraint (\ref{Converted_to_real_1_3}) is also
satisfied and, thus, $\mathbf a  = \sqrt{M}\mathbf b_l$ is a
feasible point for
(\ref{Converted_to_real_1_1})--(\ref{Converted_to_real_1_3}) that
completes the proof. \hfill$\square$

\subsection*{Proof of Theorem 2}
Let $\mathbf A^*$ be the optimal minimizer of the relaxed problem
(\ref{SDR_2_1})--(\ref{SDR_2_4}), and its rank be $r$. Consider
the following decomposition of $\mathbf A^*$
\begin{equation}
\mathbf A^* = \mathbf Y \mathbf Y ^{H}
\end{equation}
where $\mathbf Y$ is an $N \times r$ complex valued matrix. It is
trivial that if the rank of the optimal minimizer $\mathbf A^*$ of
the relaxed problem (\ref{SDR_2_1})--(\ref{SDR_2_4}) equals one,
then $\mathbf Y$ is the optimal minimizer of the original problem
(\ref{Converted_to_real_1_1})--(\ref{Converted_to_real_1_3}).
Thus, it is assumed in the following that $r > 1$.

We start by considering the following auxiliary optimization
problem
\begin{eqnarray}
\min_{\mathbf G} && Tr ( \hat{\mathbf R}^{-1} \mathbf Y \mathbf G
\mathbf Y^{H}) \label{SDR_aux_1} \\
{\rm subject \ to} && Tr( \mathbf Y \mathbf G \mathbf Y ^{H}) = M
\label{SDR_aux_2} \\
&& Tr ( \tilde{\mathbf C} \mathbf Y \mathbf G \mathbf Y ^{H} ) =
Tr ( \tilde{\mathbf C} \mathbf A^*) \label{SDR_aux_3} \\
&& \mathbf G \succeq 0 \label{SDR_aux_4}
\end{eqnarray}
where $\mathbf G$ is an $r \times r$ Hermitian matrix. The matrix
$\mathbf A$ in (\ref{SDR_2_1})--(\ref{SDR_2_4}) can be expressed
as a function of the matrix $\mathbf G$ in
\eqref{SDR_aux_1}--\eqref{SDR_aux_4} as $\mathbf A( \mathbf G) =
\mathbf Y \mathbf G \mathbf Y^{H}$. Moreover, it can be easily
shown that if $\mathbf G$ is a positive semi-definite matrix, then
$\mathbf A (\mathbf G)$ is also a positive semi-definite matrix.
In addition, if $\mathbf G$ is a feasible solution of
(\ref{SDR_aux_1})--(\ref{SDR_aux_4}), $\mathbf A (\mathbf G)$ is
also a feasible solution of (\ref{SDR_2_1})--(\ref{SDR_2_4}). The
latter is true because $\mathbf A ( \mathbf G)$ is a positive
semi-definite matrix and it satisfies the constraints $Tr (
\mathbf A( \mathbf G)) = M$ and $Tr ( \tilde{\mathbf C} \, \mathbf
A ( \mathbf G) ) = Tr( \tilde{\mathbf C} \mathbf A^*) \leq
\Delta_0$. This implies that the minimum value of the problem
(\ref{SDR_aux_1})--(\ref{SDR_aux_4}) is greater than or equal to
the minimum value of the problem (\ref{SDR_2_1})--(\ref{SDR_2_4}).

It is then easy to verify that $\mathbf G^* = \mathbf I_{r}$ is a
feasible point of the auxiliary optimization problem
(\ref{SDR_aux_1})--(\ref{SDR_aux_4}). Moreover, the fact that $ Tr
( \hat{\mathbf R}^{-1} \mathbf Y \mathbf G^* \mathbf Y^{H})= Tr (
\hat{\mathbf R}^{-1} \mathbf A^*) \triangleq \beta$ (here $\beta$
denotes the minimum value of the relaxed problem
(\ref{SDR_2_1})--(\ref{SDR_2_4})) together with the fact that the
minimum value of the auxiliary problem
(\ref{SDR_aux_1})--(\ref{SDR_aux_4}) is greater than or equal to
$\beta$, implies that $\mathbf G^* = \mathbf I_{r}$ is the optimal
solution of the auxiliary problem
(\ref{SDR_aux_1})--(\ref{SDR_aux_4}).

Next, we show that if $\mathbf G^{\prime}$ is a feasible solution
of (\ref{SDR_aux_1})--(\ref{SDR_aux_4}), then it is also an
optimum minimizer of (\ref{SDR_aux_1})--(\ref{SDR_aux_4}).
Therefore, $\mathbf A(\mathbf G^{\prime}) = \mathbf Y \mathbf
G^{\prime} \mathbf Y^H$ is also an optimum minimizer of
(\ref{SDR_2_1})--(\ref{SDR_2_4}). Towards this end, let us
consider the following dual to
(\ref{SDR_aux_1})--(\ref{SDR_aux_4}) problem
\begin{eqnarray}
\max_{\mathbf \nu_1, \nu_2, \mathbf Z } && \nu_1 M +
\nu_2 Tr( \tilde{\mathbf C} \mathbf A^*) \label{eqdual1} \\
{\rm subject \ to} && \mathbf Y^H \hat{\mathbf R}^{-1} \mathbf Y -
\nu_1 \mathbf Y^H \mathbf Y - \nu_2 \mathbf Y^H \tilde{\mathbf C}
\mathbf Y \succeq \mathbf Z \label{eqdual3} \\
&& \mathbf Z \succeq 0 \label{eqdual4}
\end{eqnarray}
where $\nu_1$ and $\nu_2$ are the Lagrange multipliers associated
with the constraints (\ref{SDR_aux_2}) and (\ref{SDR_aux_3}),
respectively, and $\mathbf Z$ is an $r \times r$ Hermitian matrix.
Note that the optimization problem
(\ref{SDR_aux_1})--(\ref{SDR_aux_4}) is convex. Moreover, it
satisfies the Slater's conditions because, as it was mentioned,
the positive definite matrix $\mathbf G = \mathbf I_{r}$ is a
feasible point for (\ref{SDR_aux_1})--(\ref{SDR_aux_4}). Thus, the
strong duality between (\ref{SDR_aux_1})--(\ref{SDR_aux_4}) and
\eqref{eqdual1}--\eqref{eqdual4} holds.

Let $\nu_1^*, \ \nu_2^*$, and $ \mathbf Z^*$ be one possible
optimal solution of the dual problem
\eqref{eqdual1}--\eqref{eqdual4}. Since strong duality holds,
$\nu_1^* M + \nu_2^* Tr( \tilde{\mathbf C} \mathbf A^*) = \beta$
and $\mathbf I_r$ is an optimal solution of the primal problem
(\ref{SDR_aux_1})--(\ref{SDR_aux_4}). Moreover, the complementary
slackness condition implies that
\begin{eqnarray} \label{compslac}
Tr(\mathbf G^* \mathbf Z^*) = Tr( \mathbf Z^*) = 0 .
\end{eqnarray}
Since $\mathbf Z^*$ has to be a positive semi-definite matrix, the
condition \eqref{compslac} implies that $\mathbf Z^* = 0$. Then it
follows from \eqref{eqdual3} that
\begin{eqnarray} \label{constmod1}
\mathbf Y^H \hat{\mathbf R}^{-1} \mathbf Y \succeq \nu_1^* \mathbf
Y^H \mathbf Y + \nu_2^* \mathbf Y^H \tilde{\mathbf C} \mathbf Y .
\end{eqnarray}

Using the fact that $\mathbf Y^H \mathbf Y$ and $ \mathbf Y^H
\tilde{\mathbf C} \mathbf Y$ are positive semi-definite matrices,
it can be easily verified that the constraint \eqref{constmod1} is
active, i.e., it is satisfied as equality for optimal $\nu_1^*$
and $\nu_2^*$. Therefore, we can write that
\begin{eqnarray}
\mathbf Y^H \hat{\mathbf R}^{-1} \mathbf Y =  \nu_1^* \mathbf Y^H
\mathbf Y +\nu_2^* \mathbf Y^H \tilde{\mathbf C} \mathbf Y .
\label{std}
\end{eqnarray}

Let $\mathbf G^\prime$ be another feasible solution of
(\ref{SDR_aux_1})--(\ref{SDR_aux_4}) different from $\mathbf I_r$. Then
the following conditions must hold
\begin{eqnarray}
&& Tr( \mathbf Y ^{H}  \mathbf Y \mathbf G^\prime) = M \\
&&Tr(\mathbf Y ^{H} \tilde{\mathbf C} \mathbf Y \mathbf G^\prime )
= Tr( \tilde{\mathbf C} \mathbf A^*) \\
&& \mathbf G^\prime \succeq 0 .
\end{eqnarray}
Multiplying both sides of the equation (\ref{std}) by $\mathbf
G^\prime$, we obtain
\begin{eqnarray} \label{multipliedG2}
\mathbf Y^H \hat{\mathbf R}^{-1} \mathbf Y  \mathbf G^\prime =
\nu_1^* \mathbf Y^H \mathbf Y \mathbf G^\prime + \nu_2^* \mathbf
Y^H \tilde{\mathbf C} \mathbf Y \mathbf G^\prime .
\end{eqnarray}
Moreover, taking the trace of the right hand and left hand sides
of \eqref{multipliedG2}, we have
\begin{eqnarray}
Tr(\mathbf Y^H \hat{\mathbf R}^{-1} \mathbf Y \mathbf G^\prime)
\!\!\!&=&\!\!\!  \nu_1^* Tr( \mathbf Y^H \mathbf Y \mathbf
G^\prime) + \nu_2^* Tr( \mathbf Y^H \tilde{\mathbf C} \mathbf Y
\mathbf G^\prime) \nonumber \\
\!\!\!&=&\!\!\!  \nu_1^* M  +\nu_2^* Tr( \tilde{\mathbf C} \mathbf
A^*) = \beta \label{std_fin-a} .
\end{eqnarray}
This implies that $\mathbf G^\prime$ is also an optimal solution
of the problem (\ref{SDR_aux_1})--(\ref{SDR_aux_4}). Therefore,
every feasible solution of (\ref{SDR_aux_1})--(\ref{SDR_aux_4}) is
also an optimal solution.

Finally, we show that there exists a feasible solution of
(\ref{SDR_aux_1})--(\ref{SDR_aux_4}) whose rank is one. As it has
been proved above, such feasible solution will also be optimal.
Let $\mathbf G = \mathbf v \mathbf v^H$. Thus, we are interested
in finding such $\mathbf v$ that
\begin{eqnarray}
&& Tr( \mathbf Y^{H}  \mathbf Y \mathbf v \mathbf v^H) = M
\label{cond82} \\
&& Tr(\mathbf Y^{H} \tilde{\mathbf C}\mathbf Y \mathbf v \mathbf
v^H ) = Tr( \tilde{\mathbf C} \mathbf A^*) . \label{cond83}
\end{eqnarray}
Equivalently, the conditions \eqref{cond82} and \eqref{cond83} can
be rewritten as
\begin{eqnarray}
&& \mathbf v^H \mathbf Y ^{H}  \mathbf Y \mathbf v = M \label{eq94} \\
&&\mathbf v^H \mathbf Y ^{H} \tilde{\mathbf C} \mathbf Y \mathbf v
= Tr( \tilde{\mathbf C} \mathbf A^*) . \label{eq95}
\end{eqnarray}
We can further write that
\begin{eqnarray}
&& \frac{1}{M} \mathbf v^H \mathbf Y^{H} \mathbf Y \mathbf v = 1
\label{eq90}\\
&& \mathbf v^H \frac{\mathbf Y ^{H} \tilde{\mathbf C} \mathbf
Y}{Tr( \tilde{\mathbf C} \mathbf A^*)} \mathbf v = 1 .
\label{eq91}
\end{eqnarray}
Moreover, equating the left hand side of \eqref{eq90} to the left
hand side of \eqref{eq91}, we obtain that
\begin{eqnarray}
\frac{1}{M} \mathbf v^H \mathbf Y^{H} \mathbf Y \mathbf v =
\mathbf v^H \frac{\mathbf Y^{H}  \tilde{\mathbf C} \mathbf Y} {Tr
( \tilde{\mathbf C} \mathbf A^* )} \mathbf v . \label{eq92}
\end{eqnarray}
Subtracting the left hand side of \eqref{eq92} from its right
hand side, we also obtain that
\begin{eqnarray}
\mathbf v^H \left( \frac{1}{M} \mathbf Y ^{H} \mathbf Y -
\frac{\mathbf Y^{H} \tilde{\mathbf C} \mathbf Y}{Tr (
\tilde{\mathbf C} \mathbf A^*)} \right) \mathbf v = \mathbf v^H
\mathbf D \mathbf v = 0 .
\end{eqnarray}
Considering the fact that $Tr( \mathbf D ) = 0$, the vector
$\mathbf v$ can be chosen as the sum of the eigenvectors of the
matrix $\mathbf D$ in \eqref{OptSol2}. Note that $\mathbf v$ can
be chosen proportional to the sum of the eigenvectors of the
matrix $\mathbf D$ such that $ \mathbf v^H \mathbf Y^{H} \mathbf Y
\mathbf v = M$ is satisfied. It will also imply that $\mathbf v^H
{\mathbf Y ^{H} \tilde{\mathbf C} \mathbf Y} \mathbf v= Tr(
\tilde{\mathbf C} \mathbf A^*)$ and, thus, \eqref{eq94} and
\eqref{eq95} are satisfied.

So far we have found a rank one solution for the auxiliary
optimization problem (\ref{SDR_aux_2})--(\ref{SDR_aux_4}), that
is, $\mathbf G = \mathbf v \mathbf v^H$. Since $\mathbf G =
\mathbf v \mathbf v^H$ is the optimal solution of the auxiliary
problem (\ref{SDR_aux_2})--(\ref{SDR_aux_4}), then $\mathbf Y
\mathbf G \mathbf Y^H = (\mathbf Y \mathbf v) (\mathbf Y \mathbf
v)^H$ is the optimal solution of the relaxed problem
(\ref{SDR_2_1})--(\ref{SDR_2_4}). Moreover, since the solution
$(\mathbf Y \mathbf v) (\mathbf Y \mathbf v)^H $ is rank-one,
$\mathbf Y \mathbf v$ is the optimal solution of the original
optimization problem
(\ref{Converted_to_real_1_1})--(\ref{Converted_to_real_1_3}). This
completes the proof. $\hfill\square$

\subsection*{\centering{Proof of Theorem 3}}
Let $\mathbf A^*$ be one optimal solution of the problem
(\ref{SDR_2_1})--(\ref{SDR_2_4}) whose rank $r$ is greater than
one. Using the rank-one decomposition of Hermitian matrices
\cite{decompostion}, the matrix $\mathbf A^*$ can be written as
\begin{equation}
\mathbf A^* = \sum_{j=1}^{r}\mathbf z_j \mathbf z_j^H
\label{decom_0}
\end{equation}
where
\begin{eqnarray}
&& \mathbf z_j ^H \mathbf z_j = \frac{1}{r} Tr ( \mathbf A^*) =
\frac{M}{r} , \quad j=1, \hdots, r \label{decom_1} \\
&& \mathbf z_j ^H \tilde{\mathbf C} \mathbf z_j = \frac{1}{r} Tr (
\tilde{\mathbf C} \mathbf A^*) , \quad j=1, \hdots, r .
\label{decom_2}
\end{eqnarray}

Let us show that the terms $\mathbf z_j^H \hat{\mathbf R}^{-1}
\mathbf z_j$, $j=1, \hdots, r$ are equal to each other for all
$j=1, \hdots, r$. We prove it by contradiction assuming first that
there exist such $\mathbf z_m$ and $\mathbf z_n$, $m \neq n$ that
$\mathbf z_m^H \hat{\mathbf R}^{-1} \mathbf z_m < \mathbf z_n^H
\hat{\mathbf R}^{-1} \mathbf z_n$. Let the matrix $\mathbf A^*_0$
be constructed as $\mathbf A^*_0 =\mathbf A^* - \mathbf z_n
\mathbf z_n^H + \mathbf z_m \mathbf z_m^H$. It is easy to see that
$Tr ( \mathbf A^* ) = Tr ( \mathbf A^*_0 )$ and $Tr (
\tilde{\mathbf C} \mathbf A^* ) = Tr ( \tilde{\mathbf C} \mathbf
A^*_0)$, which means that $\mathbf A^*_0$ is also a feasible
solution of the problem (\ref{SDR_2_1})--(\ref{SDR_2_4}). However,
based on our assumption that $\mathbf z_m^H \hat{\mathbf R}^{-1}
\mathbf z_m < \mathbf z_n^H \hat{\mathbf R}^{-1} \mathbf z_n$, it
can be concluded that $Tr ( \hat{\mathbf R}^{-1} \mathbf A^*_0) <
Tr ( \hat{\mathbf R}^{-1} \mathbf A^*_0)$ that is obviously a
contradiction. Thus, all terms $\mathbf z_j^H \hat{\mathbf R}^{-1}
\mathbf z_j$, $j = 1, \hdots, r$ must take the same value. Using
this fact together with the equations (\ref{decom_1}) and
(\ref{decom_2}), we can conclude that $r \mathbf z_j \mathbf
z_j^H$ for any $j =1, \hdots, r$ is the optimal solution of the
relaxed optimization problem (\ref{SDR_2_1})--(\ref{SDR_2_4})
which has rank one. Thus, the optimal solution of the original
problem
(\ref{Converted_to_real_1_1})--(\ref{Converted_to_real_1_3}) is
$\sqrt{r} \mathbf z_j$ for any $j = 1, \hdots, r$. Since the
vectors $\mathbf z_j$, $j=1,..., r$ in \eqref{decom_0} are
linearly independent and each of them gives an optimal solution to
the problem
(\ref{Converted_to_real_1_1})--(\ref{Converted_to_real_1_3}), we
conclude that the optimal solution to
(\ref{Converted_to_real_1_1})--(\ref{Converted_to_real_1_3}) is
not unique up to a phase rotation. However, it contradicts the
assumption that the optimal solution of
(\ref{Converted_to_real_1_1})--(\ref{Converted_to_real_1_3}) is
unique up to a phase rotation. Thus, the optimal solution $\mathbf
A^*$ to the relaxed problem (\ref{SDR_2_1})--(\ref{SDR_2_4}) must
be rank-one. This completes the proof. $\hfill\square$

\newpage

\begin{thebibliography}{1}

\bibitem{Trees} H.~L.~Van~Trees, {\it Optimum Array Processing}.
New York: Wiley, 2002.

\bibitem{RMB} I.~S.~Reed, J.~D.~Mallett, and L.~E.~Brennan, ``Rapid
convergence rate in adaptive arrays,'' {\it IEEE Trans. Aerosp.
Electron. Syst.}, vol.~10, pp.~853–-863, Nov.~1974.

\bibitem{Griffits} L.~J.~Griffiths and C.~W.~Jim, ``An alternative
approach to linearly constrained adaptive beamforming,'' {\it IEEE
Trans. Antennas Propagat.}, vol.~30, pp.~27–-34, Jan.~1982.

\bibitem{Hung} E.~K.~Hung and R.~M.~Turner, ``A fast beamforming
algorithm for large arrays,'' {\it IEEE Trans. Aerosp. Electron.
Syst.}, vol.~19, pp.~598–-607, July~1983.

\bibitem{Gershman2} A.~B.~Gershman, ``Robust adaptive beamforming
in sensor arrays,'' {\it Int. J. Electron. Commun.}, vol.~53,
pp.~305–-314, Dec.~1999.

\bibitem{Cox} H.~Cox, R.~M.~Zeskind, and M.~H.~Owen, ``Robust adaptive
beamforming,'' {\it IEEE Trans. Acoust., Speech, Signal
Processing}, vol.~ASSP-35, pp.~1365–-1376, Oct.~1987.


\bibitem{Abramovich} Y.~I.~Abramovich, ``Controlled method for
adaptive optimization of filters using the criterion of maximum
SNR,'' {\it Radio Eng. Electron. Phys.}, vol.~26, pp.~87–-95,
Mar.~1981.

\bibitem{Feldman} D.~D.~Feldman and L.~J.~Griffiths, ``A projection
approach to robust adaptive beamforming,'' {\it IEEE Trans. Signal
Processing}, vol.~42, pp.~867-–876, Apr.~1994.

\bibitem{Chang} L.~Chang and C.~C.~Yeh, ``Performance of DMI and
eigenspace-based beamformers,'' {\it IEEE Trans. Antennas
Propagat.}, vol.~40, pp.~1336–-1347, Nov.~1992.


\bibitem{Scharf} J.~K.~Thomas, L.~L.~Scharf, and D.~W.~Tufts, ``The
probability of a subspace swap in the SVD,'' {\it IEEE Trans.
Signal Processing}, vol.~43, pp.~730–-736, Mar.~1995.

\bibitem{Vorobyov1} S.~A.~Vorobyov, A.~B.~Gershman, Z.-Q.~Luo,``Robust
adaptive beamforming using worst-case performance optimization: A
solution to the signal mismatch problem,'' {\it IEEE Trans. Signal
Processing}, vol.~51, pp.~313--324, Feb.~2003.

\bibitem{Vorobyov2} S.~A.~Vorobyov, A.~B.~Gershman, Z.-Q.~Luo, and
N.~Ma, ``Adaptive beamforming with joint robustness against
mismatched signal steering vector and interference
nonstationarity,'' {\it IEEE Signal Processing Lett.}, vol.~11,
pp.~108-–111, Feb.~2004.

\bibitem{LiStoica} J.~Li, P.~Stoica, and Z.~Wang, ``On robust Capon
beamforming and diagonal loading,'' {\it IEEE Trans. Signal
Process.}, vol.~51, pp.~1702-–1715, July~2003.

\bibitem{Lorenz} R.~G.~Lorenz and S.~P.~Boyd, ``Robust minimum variance
beamforming,'' {\it IEEE Trans. Signal Process.}, vol.~53,
pp.~1684-–1696, May~2005.

\bibitem{Shahbaz} S.~Shahbazpanahi, A.~B.~Gershman, Z.-Q.~Luo, and
K.~M.~Wong, ``Robust adaptive beamforming for general-rank signal
models,'' {\it IEEE Trans. Signal Process.}, vol.~51,
pp.~2257–-2269, Sep.~2003.

\bibitem{Vorobyov7} S.~A.~Vorobyov, Y.~Rong, and A.~B.~Gershman,
``Robust adaptive beamforming using probability-constrained
optimization,'' in {\it Proc. IEEE SSP Workshop}, Bordeaux,
France, July~2005, pp.~934--939.

\bibitem{Vorobyov4} S.~A.~Vorobyov, H.~Chen, and A.~B.~Gershman,
``On the relationship between robust minimum variance beamformers
with probabilistic and worst-case distrortionless response
constraints,'' {\it IEEE Trans. Signal Processing}, vol.~56,
pp.~5719--5724, Nov.~2008.

\bibitem{Vorobyov3} S.~A.~Vorobyov, A.~B.~Gershman, and Y.~Rong,
``On the relationship between the worst-case optimization-based
and probability-constrained approaches to robust adaptive
beamforming,'' in {\it Proc. IEEE ICASSP}, Honolulu, HI,
Apr.~2007, pp.~977-–980.

\bibitem{Nasr} A.~Hassanien, S.~A.~Vorobyov, and K.~M.~Wong, ``Robust
adaptive beamforming using sequential quadratic programming,'' in
{\it Proc. IEEE ICASSP}, Las Vegas, NV, Apr.~2008, pp.~2345--2348.

\bibitem{Vorobyov5} A.~Hassanien, S.~A.~Vorobyov, and K.~M.~Wong,
``Robust adaptive beamforming using sequential programming: An
iterative solution to the mismatch problem,'' {\it IEEE Signal
Processing Lett.}, vol.~15, pp.~733--736, 2008.

\bibitem{LiStoicaWang} J.~Li, P.~Stoica, and Z.~Wang, ``Doubly
constrained robust capon beamformer,'' {\it IEEE Trans. Signal
Processing}, vol.~52, pp.~2407--2423, Sept.~2004.

\bibitem{ZH} S.~Zhang and Y.~Huang, ``Complex quadratic
optimization and semidefinite programming,'' {\it SIAM J. Optim.},
vol.~16, no.~3, pp.~871--890, 2006.

\bibitem{Luo2010} Z.-Q.~Luo, W.-K.~Ma, A.~M.-C.~So, Y.~Ye, and
S.~Zhang, ``Semidefinite relaxation of quadratic optimization
problems,'' {\it IEEE Signal Processing Magazin}, vol.~27, no.~3,
pp.~20--34, May~2010.


\bibitem{Nesterov} Y.~S.~Nesterov, ``Semidefinite relaxation and
nonconvex quadratic optimization,'' {\it Optim. Methods Softw.},
vol.~9, no.~1--3, pp.~141--160, 1998.

\bibitem{PhanVor} K.~T.~Phan, S.~A.~Vorobyov, N.~D.~Sidiropoulos,
and C.~Tellambura, ``Spectrum sharing in wireless networks via
QoS-aware secondary multicast beamforming,'' {\it IEEE Trans.
Signal Processing}, vol.~57, pp.2323--2335, June~2009.

\bibitem{Zhang} S. Zhang, ``Quadratic maximization and semidefinite
relaxation,'' {\it Math. Program. A}, vol.~87, pp.~453–-465, 2000.

\bibitem{Tom} Z.-Q.~Luo, N.~D.~Sidiropoulos, P.~Tseng, and
S.~Zhang, ``Approximation bounds for quadratic optimization with
homogeneous quadratic constraints,'' {\it SIAM J. Optim.},
vol.~18, no.~1, pp.~1--28, Feb.~2007.

\bibitem{Sedumi} J.~F.~Sturm, ``Using SeDuMi 1.02,
a MATLAB toolbox for optimization over symmetric cones,'' {\it
Optim. Meth. Softw.}, vol.~11–12, pp.~625–-653, Aug.~1999.
Available [Online]: http://sedumi.ie.lehigh.edu/

\bibitem{Shahbaz2} A.~B.~Gershman, Z.-Q.~Luo, and
S.~Shahbazpanahi, ``Robust adaptive beamforming based on
worst-case performance optimization,'' in {\it Robust Adaptive
Beamforming}, P. Stoica and J. Li, Eds. Hoboken, NJ: Wiley, 2006,
pp.~49–-89.

\bibitem{Gershman} L.~Lei, J.~P.~Lie, A.~B.~Gershman, and C.~M.~S.~See,
``Robust adaptive beamforming in partly calibrated sparse sensor
arrays,`` {\it IEEE Trans. Signal Processing}, vol.~58,
pp.~1661--1667, Mar.~2010.

\bibitem{beck}
A.~Beck and Y.~C.~Eldar, ``Strong duality in nonconvex quadratic
optimization with two quadratic constraints,'' {\it SIAM J.
Optimization}, vol.~17, no.~3, pp.~844--860, 2006.

\bibitem{beck2}
A.~Beck and Y.~C.~Eldar, ``Doubly constrained robust Capon
beamformer with ellipsoidal uncertainty sets,'' {\it IEEE Trans.
Signal Processing}, vol.~55, pp.~753--758, Feb.~2007.


\bibitem{cvx} Available [Online]: http://cvxr.com/cvx/

\bibitem{coherentscattering}
J. Goldberg and H. Messer, ``Inherent limitations in the
localization of a coherently scattered source,'' {\it IEEE Trans.
Signal Processing}, vol.~46, pp.~3441–-3444, Dec.~1998.


\bibitem{decompostion}
Y.~Huang and S.~Zhang, ``Complex Matrix Decomposition and
Quadratic Programming,'' {\it Mathematics of Operations Research},
vol.~32, no.~3, pp.~758--768, Aug.~2007.
\end{thebibliography}
\end{document}